\documentclass[letterpaper]{article} % DO NOT CHANGE THIS
\usepackage{aaai23}  % [submission]
\usepackage{times}  % DO NOT CHANGE THIS
\usepackage{helvet}  % DO NOT CHANGE THIS
\usepackage{courier}  % DO NOT CHANGE THIS
\usepackage[hyphens]{url}  % DO NOT CHANGE THIS
\usepackage{graphicx} % DO NOT CHANGE THIS
\usepackage{natbib}  % DO NOT CHANGE THIS AND DO NOT ADD ANY OPTIONS TO IT
\usepackage{caption} % DO NOT CHANGE THIS AND DO NOT ADD ANY OPTIONS TO IT
\usepackage{multirow}
\usepackage[utf8]{inputenc} % allow utf-8 input
\usepackage[T1]{fontenc}    % use 8-bit T1 fonts
\usepackage{booktabs}       % professional-quality tables
\usepackage{amsfonts}       % blackboard math symbols
\usepackage{nicefrac}       % compact symbols for 1/2, etc.
\usepackage{microtype}      % microtypography
\usepackage{xcolor}         % colors
\usepackage{kotex}         % colors
\usepackage{sidecap,caption}
\usepackage{appendix}
\usepackage{apxproof}
\usepackage{wrapfig}
\usepackage{enumitem}
\usepackage{float}
\usepackage{physics}
\usepackage{bm}
\usepackage{bbm}
\usepackage{textcomp}
\usepackage[caption=false]{subfig}
\usepackage[export]{adjustbox}
\usepackage{tikz}
\newcommand*\circled[1]{\tikz[baseline=(char.base)]{
            \node[shape=circle,draw,inner sep=.5pt] (char) {\small #1};}}
\usepackage{amsthm}
\newtheorem{theorem}{Theorem}
\newtheorem{definition}{Definition}

\usepackage{thmtools}
\usepackage{thm-restate}
\declaretheorem[name=Lemma]{lemma}
\newtheorem{assumption}{Assumption}

\usepackage{amssymb}
\usepackage{balance}
\usepackage{qcircuit}
\usepackage{tikz}
\usepackage{array}
\usepackage{xcolor}
\usepackage{mathtools}
\usepackage[algo2e, linesnumbered, ruled, vlined]{algorithm2e}
\usepackage{etoolbox}
\newcommand{\ppsi}{\ensuremath{\psi^+}}
\newcommand{\mpsi}{\ensuremath{\psi^-}}

\newcommand{\policy}{\ensuremath{\pi}}

\newcommand{\phib}{\ensuremath{\bm{\phi}}}
\newcommand{\thetab}{\ensuremath{\bm{\theta}}}

\newcommand{\phis}{\ensuremath{\bm{\phi}}}

\newcommand{\taus}{\ensuremath{\bm{\tau}}}

\newcommand{\BfPara}[1]{{\noindent\bf#1.}\xspace}

\usepackage[normalem]{ulem}

\DeclareMathOperator*{\argmax}{arg\,max}

\usepackage{newfloat}
\usepackage{listings}
\DeclareCaptionStyle{ruled}{labelfont=normalfont,labelsep=colon,strut=off} % DO NOT CHANGE THIS
\floatstyle{ruled}
\newfloat{listing}{tb}{lst}{}
\floatname{listing}{Listing}
\title{Quantum Multi-Agent Meta Reinforcement Learning}
\author {
    Won Joon Yun\textsuperscript{\rm 1},
    Jihong Park\textsuperscript{\rm 2}, 
    Joongheon Kim\textsuperscript{\rm 1}
}
\affiliations {
    \textsuperscript{\rm 1} School of Electrical Engineering, Korea University, Seoul, Republic of Korea\\
    \textsuperscript{\rm 2} School of Information Technology, Deakin University, Geelong, VIC, Australia\\
    ywjoon95@korea.ac.kr, 
    jihong.park@deakin.edu.au, joongheon@korea.ac.kr
}

\usepackage{bibentry}
\begin{document}

\maketitle

\begin{abstract}
Although quantum supremacy is yet to come, there has recently been an increasing interest in identifying the potential of quantum machine learning (QML) in the looming era of practical quantum computing. Motivated by this, in this article we re-design multi-agent reinforcement learning (MARL) based on the unique characteristics of quantum neural networks (QNNs) having two separate dimensions of trainable parameters: \emph{angle} parameters affecting the output qubit states, and \emph{pole} parameters associated with the output measurement basis. Exploiting this dyadic trainability as meta-learning capability, we propose \emph{quantum meta MARL (QM2ARL)} that first applies angle training for meta-QNN learning, followed by pole training for few-shot or local-QNN training. To avoid overfitting, we develop an \emph{angle-to-pole regularization} technique injecting noise into the pole domain during angle training. Furthermore, by exploiting the pole as the memory address of each trained QNN, we introduce the concept of \emph{pole memory} allowing one to save and load trained QNNs using only two-parameter pole values. We theoretically prove the convergence of angle training under the angle-to-pole regularization, and by simulation corroborate the effectiveness of QM2ARL in achieving high reward and fast convergence, as well as of the pole memory in fast adaptation to a time-varying environment.
\end{abstract}

\section{Introduction}\label{sec:1}

Spurred by recent advances in quantum computing hardware and machine learning (ML) algorithms, quantum machine learning (QML) is closer than ever imagined. The noisy intermediate-scale quantum (NISQ) era has already been ushered in, where quantum computers run with up to a few hundred qubits \cite{cho2020ibm}. Like the neural network (NN) of classical ML, the parameterized quantum circuit (PQC), also known as a quantum NN (QNN), has recently been introduced as the standard architecture for QML \cite{chen20,jerbi2021variational,lockwood2020reinforcement}. According to IBM's roadmap, it is envisaged to reach the full potential of QML by around 2026 when quantum computers can run with 100k qubits \cite{roadmap2022}. 

Motivated by this trend, recent works have started re-implementing existing ML applications using QNNs, ranging from image classification to reinforcement learning (RL) tasks \cite{schuld2020circuit,chen20}. Compared to classical ML, QML is still in its infancy and too early to demonstrate quantum supremacy in accuracy and scalability due to the currently limited number of qubits. Instead, the main focus in the current research direction is to identify possible challenges and novel potential in QNN-based QML applications \cite{Schuld2022QML}.

Following this direction, in this article we aim to re-design multi-agent reinforcement learning (MARL) using QML, i.e., \emph{quantum MARL (QMARL)}. The key new element is to leverage the novel aspect of the QNN architecture, having two separate dimensions of trainable parameters. As analogous to training a classical NN by adjusting weight parameters, the standard way of training a QNN is optimizing its circuit parameters, or equivalently the rotation of the \emph{angle~$\phib$} of the parameter quantum circuit's output qubit states that are represented on the surface of the Bloch sphere \cite{bloch1946nuclear}. Unlike classical ML, the QNN's output for the loss calculation is not deterministic, but is measurable after projecting multiple observations with a projector on a Hilbert space \cite{nielsen2002quantum}. According to the quantum kernel theory \cite{schuld2019quantum}, this projector is represented as the \emph{pole $\thetab$} of the Bloch sphere, and is tunable by rotating the pole, providing another dimension for QNN training.

\begin{figure*}[t!]
\centering
   \includegraphics[width=0.83\textwidth]{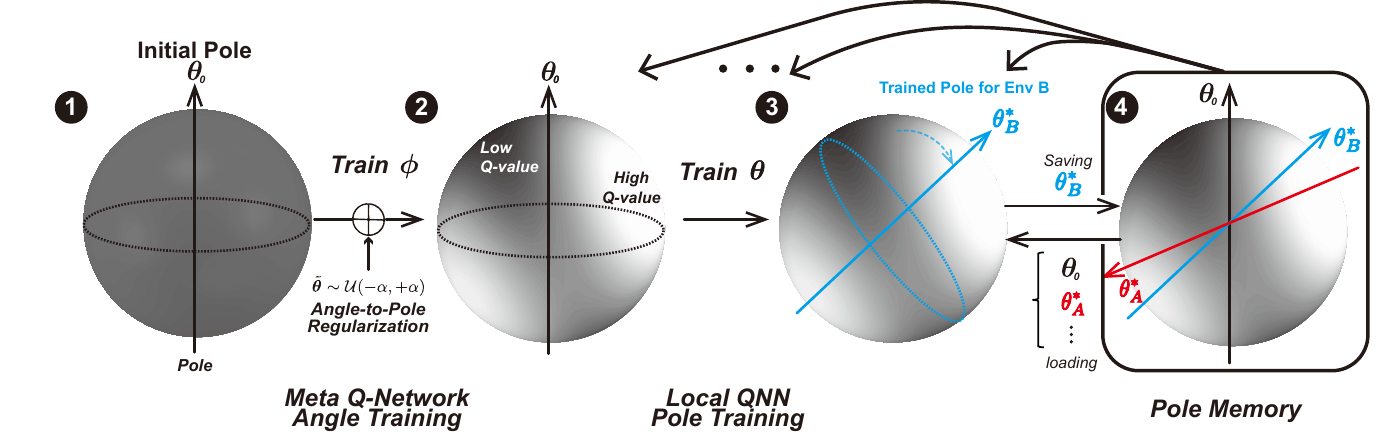}
% \vspace{-3mm}
\caption{The key concept of quantum multi-agent meta reinforcement learning for a single qubit. A classical data (\textit{e.g.}, state or observation) with the \textit{Q}-value is mapped onto the Bloch sphere. The lighter and the darker surfaces stand for the high and low \textit{Q}-value, respectively.
The first Bloch sphere presents the untrained \textit{Q}-value map. From meta-QNN angle training with the noise regularizer, the parameter quantum circuit's parameters $\phis$ are trained, and the output is presented as the second Bloch sphere. Then, the pole parameters $\thetab$ are trained. 
If catastrophic remembering for adapting to multiple environments is needed, the position of the pole is initialized to a certain pole (\textit{e.g.,} an initial pole or optimal pole of other environments).}\label{fig:abstract}
\end{figure*}

Interpreting such dyadic QNN trainability as meta-learning capability \cite{finn2017model}, we propose a novel QMARL framework coined \emph{quantum meta MARL (QM2ARL)}. As Fig.~\ref{fig:abstract} visualizes, QM2ARL first trains angle~$\phib$ for meta Q-network learning, followed by training pole $\thetab$ for few-shot or local training. The latter step is much faster in that $\thetab$ has only two parameters (i.e., polar and azimuthal angles in spherical coordinates\typeout{, (See~Fig.~1 in Supplementary Materials)}. When the number of environments is limited, angle training may incur overfitting, making the meta Q-network ill-posed. To avoid this while guaranteeing convergence, we develop an \emph{angle-to-pole regularization} method that injects random noise into the pole domain during angle training, and theoretically proves the convergence of angle training under the presence of angle-to-pole regularization. 

Furthermore, it is remarkable that each locally-trained QNN in QM2ARL can be uniquely represented only using its pole deviation $\thetab$ from the origin or equivalently the meta-trained QNN. Inspired by this, we introduce the concept of \emph{pole memory} that can flexibly save and load the meta-trained and locally-trained QNNs only using their pole values as the memory addresses. Thanks to the two-parameter space of $\thetab$, the pole memory can store $K$ different QNN configurations only using $2K$ parameters, regardless of the QNN sizes, i.e., $\phib$ dimensions. By simulation, we show the effectiveness of pole memory to cope with time-varying and cyclic environments wherein QM2ARL can swiftly adapt to a revisiting environment by loading the previous training history.

\BfPara{Contributions} The main contribution of this work is summarized as follows. First, we propose QM2ARL, the first QMARL framework that utilizes both the angle and pole domains of a QNN for meta RL. Second, we develop the angle-to-pole regularization technique to avoid the meta-trained QNN's overfitting. Third, we theoretically prove the bounded convergence of meta-QNN training in the presence of the angle-to-pole regularization, which is non-trivial as the gradient variance during angle training may diverge. Fourth, we introduce the pole memory, and show its effectiveness in fast adaptation to a time-varying environment. Lastly, by simulation we validate that QM2ARL achieves higher rewards and faster convergence than several baselines, including the na\"ive CTDE-QMARL~\cite{yun2022QMARL} under the MARL environments of a two-step game~\cite{SunehagLGCZJLSL18,RashidSWFFW20,SonKKHY19} and a single-hop offloading scenario~\cite{yun2022QMARL}.

\section{Preliminaries of QMARL}
\subsection{QMARL Setup}
\BfPara{Notation} The bold symbol in this paper denotes the vectorized form of the normal symbol. $\ket{\psi}$, $\phis \triangleq \{\phi_1, \cdots, \phi_k, \cdots, \phi_\abs{\phis}\}$, and $\thetab \triangleq \{\theta_1, \cdots, \theta_k, \cdots, \theta_\abs{\thetab}\}$ are defined as an entangled quantum state, the parameters of parameter quantum circuit, and the parameters of measurement, respectively. Here, $\phi_k$ and $\theta_k$ are the $k$-th entries of $\phis$ and $\thetab$. Moreover, $\otimes$, $(\cdot)^\dagger$, and $(\cdot)^T$ denote Kronecker product, complex conjugate operator, and transpose operator, respectively. The terms \textit{``observation"} and \textit{``observable"} are used in this paper where the \textit{``observation"} is information that an agent locally obtained in a multi-agent environment, whereas the \textit{``observable"} is an operator whose property of the quantum state is measured. 

\BfPara{Multi-agent Settings}
QMARL is modeled with \textit{decentralized partially observable Markov decision process (DecPOMDP)}~\cite{Springer2016_POMDP}.
In DecPOMDP, the local observation, true state on current/next time-step, action, reward, and the number of agents are denoted as $o \in \mathbb{R}^\abs{o}$, $s,s' \in \mathbb{R}^\abs{s}$, $a \in \mathbb{R}^\abs{a}$, $r(s,\mathbf{a})$, and $N$ respectively. The joint observation or joint action is denoted as $\mathbf{o} = \{o^n\}_{n=1}^N$, and $\mathbf{a}=\{a\}^N_{n=1}$, respectively. 
Each agent has (i) QNN-based policy for execution and (ii) \textit{Q}-network and target \textit{Q}-network for training.

\begin{figure}[t!]
\centering
\includegraphics[width=1\columnwidth]{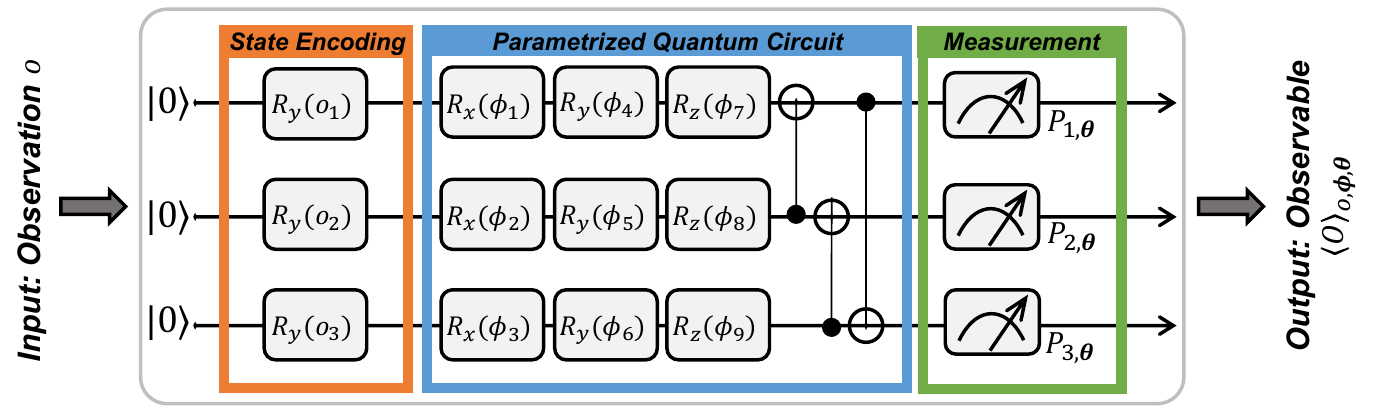}
    \caption{QNN architecture where the PQC and the measurement have tunable parameters $\phi_k$ (angle) and $\theta$ (pole), respectively.}\label{fig:qmarl}
% \vspace{-3mm}
\end{figure}  
\subsection{Quantum Neural Network}
As illustrated in Fig.~\ref{fig:qmarl}, QNN consists of three components: state encoding, PQC, and measurement~\cite{killoran2019continuous}, as elaborated next.
\typeout{Note that the basics of quantum computing notations and concepts are summarized in Appendix. }

\BfPara{State Encoding}
The state encoder feeds the observation $o$ into the circuit. 
In one-variable state encoding, the encoding process is written as  $\ket{\psi_{o}}= \otimes^{L}_{k=1} [R_y(o_k)\ket{0}]$, where $o_k$ and $R_y$ stand for the $k$-th entry of observation and the rotation gate over the $y$-axis, respectively.
As an extension to our QNN architecture, four-variable state encoding is presented for feeding state information into the QNN-based state-value network in QMARL~\cite{yun2022QMARL}.
According to \cite{lockwood2020reinforcement}, the state encoding with fewer qubit variables (\textit{e.g.,} one- or two-variable state encoding) will not suffer from performance degradation. In addition, a classical neural encoder is utilized to prevent dimensional reduction~\cite{lockwood2021playing}. This paper uses one-variable state encoding, well-known and widely used in QML.

\BfPara{Parameterized Quantum Circuit}
A QNN is designed to emulate the computational procedure of NNs. 
QNN takes encoded quantum state $\ket{\psi_o}$, which is encoded by the state encoder. Then, PQC consists of unitary gates such as rotation gates (\textit{i.e.,} $R_x$, $R_y$, and $R_z$) and \textit{controlled-NOT (CNOT)} gates. Each rotation gate has its trainable parameter $\phi$, and the rotation gate transforms and entangles the probability amplitude of the quantum state. This process is expressed with a unitary operator $U(\phis)$, written as $\ket{\psi_{o,\phis}}=U(\phis)\ket{\psi_{o}}$. The quantum state of PQC's output $\ket{\psi_{o,\phis}}$ is mapped into a $2^L$-dimensional quantum space similar to NNs' feature/latent space. This paper utilizes a basic operator block as three rotation gates and a CNOT gate for each qubit~\cite{o2004quantum}, which is
controlled by the adjacent neighboring qubit circularly.

\BfPara{Measurement}
A projective measurement is described by a Hermitian operator $O \in [-1,1]^{|\mathcal{M}_a|}$, called \textit{observable}. 
According to the \textit{Born rule}, its spectral decomposition defines the outcomes of this measurement as $O=\sum_{m\in \mathcal{M}_a} \alpha_m P_{m,\thetab}$, where $P_{m,\thetab}$ and $\alpha_m$ denote the orthogonal projections of the $m$-th qubit and the eigenvalue of the measured state, respectively. The orthogonal projections are written as  $P_{m,\thetab} = I^{\otimes m-1} \otimes M_{\theta_m} \otimes I^{\otimes L-m}$, where $I$ and $\mathcal{M}_a$ stand for the $2\times 2$ identity matrix and the $\abs{a}$-combination of  $\mathbb{N}[1,L]$, respectively. 
Throughout this paper, let $M_{\theta_m}=  \left[\begin{smallmatrix}\cos\theta_m &-\sin \theta_m\\ -\sin \theta_m & \cos \theta_m \end{smallmatrix} \right]$ denote the measurement operator of the m-th qubit, which is decomposed into:
$M_{\theta_m} =
\begin{bmatrix}
\cos\frac{{\theta_m}}{2} & \sin\frac{{\theta_m}}{2} \\
-\sin\frac{{\theta_m}}{2} & \cos\frac{{\theta_m}}{2}
\end{bmatrix}
\cdot
\begin{bmatrix}
1 & 0 \\
0 & -1
\end{bmatrix}
\cdot
\begin{bmatrix}
\cos\frac{{\theta_m}}{2} & -\sin\frac{{\theta_m}}{2} \\
\sin\frac{{\theta_m}}{2} & \cos\frac{{\theta_m}}{2}
\end{bmatrix}$.

The quantum state gives the outcome $\alpha_m$ and is projected onto the measured state $P_{m,\thetab} \ket{\psi_{o,\phis}}/\sqrt{p(m)}$ with probability $p(m) = \bra{\psi_{o,\phis}} P_{m,\thetab} \ket{\psi_{o,\phis}} = {\expval{P_m}}_{{o,\phis}}$. 
The expectation of the observable $O$ concerning $\ket{\psi_{o,\phis}}$ is $\mathbb{E}_{\psi_{o,\phis},\thetab}[O] = \sum_m p(m) \alpha_m = \expval{O}_{{o,\phis,\thetab}}$. 

\subsection{QNN Implementation to Reinforcement Learning}
Suppose that every MARL agent has its own \textit{Q}-network and policy. Motivated by \cite{jerbi2021variational}, we define the \textit{Q}-network as follows.
\begin{definition}[\textsc{Q-Network}]\label{def1}
Given QNN acting on $L$ qubits, taking input observation $o$, the trainable PQC parameters (\textit{i.e.}, angle parameters) $\phis \in [-\pi,\pi]^\abs{\phis}$ and their corresponding unitary transformation $U(\phis)$ produces the quantum state $\ket{\psi_{o,\phis}} = U(\phis) \ket{\psi_o}$. The quantum state $\ket{\psi_{o,\phis}}$ and trainable pole parameters $\thetab \in [-\pi,\pi]^\abs{a}$, and a scaling hyperparameter $\beta \in \mathbb{R}$ make an observable with a projection matrix $P_m$ associated with an action $a$ and $m$-th qubit. The \textit{Q}-network is defined as $Q(o,a;\phis,\thetab) = \beta\expval{O_a}_{o,\phis,\thetab} =\beta\ev{\sum_{m\in\mathcal{M}_a} P_{m,\thetab}}{\psi_{o,\phis}}$. 
\end{definition} 
In QMARL architecture, the policy can be expressed with the softmax function of the \textit{Q}-network. Note that the policy proposed by \cite{jerbi2021variational} has utilized measurement as weighted Hermitian that are trained in a classical way. However, we consider that the \textit{Q}-network/policy utilizes pole parameters, which will be further discussed.

\section{Quantum Multi-Agent Meta Reinforcement Learning}
Hereafter, we use the prefix ``\textit{meta-}" for the terms that are related to meta learning, \textit{e.g.,} meta \textit{Q}-network, and meta agent.
We divide QMARL into two stages, \textit{i.e.}, meta-QNN angle training and local QNN pole training.
In this section, we first describe the meta-QNN angle training, where the angle parameters are trained in the angle domain $[-\pi,\pi]^{\abs{\phis}}$. 
Then, we present the local-QNN pole training where the pole parameters are trained in the pole domain $[-\pi,\pi]^{\abs{\thetab}}$ with the fixed meta-QNN angle parameters.
Lastly, we introduce a QM2ARL application exploiting its pole memory to cope with catastrophic forgetting.

\subsection{Meta QNN Angle Training with Angle-to-Pole Regularization}\label{sec:3.1}
Meta-QNN angle training aims to be generalized over various environments. This is, however challenged by the limited size of the Q-network and/or biased environments. Injecting noise as regularization can ameliorate this problem. One possible source of noise is the state noise \cite{wilson2021optimizing}, which is neither controllable nor noticeably large, particularly with a small number of qubits. Alternatively, during the \emph{Q}-network angle training, we consider injecting an artificial noise into the pole domain as follows.
\begin{definition}[\textsc{Angle-to-Pole Regularization}]\label{def:geometric} Let an angle noise be defined as a multivariate independent random variable, $\tilde{\thetab} \sim \mathcal{U}(-\alpha, +\alpha)$, where $\forall \alpha \in [0, \pi]$, where the probability density function $f_{\tilde{\thetab}}(x) = \frac{1}{2\alpha}$, $\forall x \in \mathcal{U}$. 
\end{definition}
The angle-to-pole regularization injects artificial noise on the pole parameters. Thus, it impacts on the projection matrix, the meta \textit{Q}-network, and finally the loss function of meta \textit{Q}-network. Since QM2ARL has an independent MARL architecture, following {independent deep \textit{Q}-networks (IDQN)} ~\cite{tampuu2017multiagent} and the \textit{double deep {Q}-networks (DDQN)} ~\cite{van2016deep}, the loss function is given as a temporal difference of the meta \textit{Q}-network: $\mathcal{L}(\phis;\thetab+\tilde{\thetab}, \mathcal{E}) = \frac{1}{n(\mathcal{E})}\sum_{\expval{o,a,r,o'} \in \mathcal{E}}[r+ Q(o',\argmax_{a'}a';\phis',\thetab) -  Q(o,a;\phis,\thetab+\tilde{\thetab})]^2$.
In contrast to the classical gradient descent, the quantum gradient descent of loss can be obtained by the \textit{parameter shift rule}~\cite{mitarai18,schuld19}\typeout{, as detailed in Appendix}.
After calculating the loss gradient, angle parameters are updated. 

Note here that while the pole parameters are not yet updated during meta Q-network angle training, the angle-to-pole regularization affects the training of angle parameters in the meta \textit{Q}-network. It is unclear whether such a random regularization impact obstructs the QM2ARL convergence, calling for convergence analysis. To this end, we focus on analyzing the convergence of the meta-QNN angle training under the existence of the angle-to-pole regularization while ignoring the measurement noise. 

In the recent literature on quantum stochastic optimization, the QNN convergence has been proved under a black-box formalization \cite{harrow2021low} and under the existence of the measurement noise with noise-free gates~\cite{gentini2020noise}. Based on this, in order to focus primarily on the impact of the angle-to-pole regularization, we assume the convergence of meta Q-network angle training without the regularization as below.

\begin{assumption}[\textsc{Convergence Without Regularization}]
Without the angle-to-pole regularization, at the $t$-th epoch, the angle training error of a meta \textit{Q}-network with its suboptimal parameter $\phis^*$ is upper bounded by a constant $\epsilon_t \geq 0$, \textit{i.e.}, $\norm{\phis_t-\phis^*} \leq \epsilon_t$.
\end{assumption}
For the sake of mathematical amenability, we additionally consider the following assumption.
\begin{assumption}[\textsc{Bi-Lipschitz Continuous}]
Meta \textit{Q}-network is \textit{bi-Lipschitz continuous}, \textit{i.e.}, $
         L_1 \| \sum_{j=t}^{\infty} \eta_j \nabla_{\phis} \mathcal{L}(\phi,\theta;\mathcal{E}_j)\| \leq \|\phis_t-\phis^* \| \leq L_2 \|\sum_{j=t}^{\infty}\eta_j \nabla_{\phis}\mathcal{L}(\phi,\theta;\mathcal{E}_j)\|$, where $L_2 \geq L_1 > 0$, $\eta_j>0$ and $\mathcal{E}_j$ are constant learning rate and an episode at epoch $j$, respectively.
\end{assumption}

Then, we are ready to show the desired convergence.
\begin{theorem}
With Assumptions 1 and 2, the angle training error of a meta \textit{Q}-network with angle-to-pole regularization at epoch $t$ is upper bounded by a constant $\mathbb{E}_{\tilde{\thetab}}\norm{\tilde\phis_t-\phis^*} \leq \frac{\sin\alpha}{\alpha} (\epsilon_t +  \epsilon_{t}'). $
% \begin{equation}
% \mathbb{E}_{\tilde{\thetab}}\norm{\tilde\phis_t-\phis^*} \leq \frac{\sin\alpha}{\alpha} (\epsilon_t +  \epsilon_{t}'). \label{Eq:Thm1}
% \end{equation}
\end{theorem}

\noindent\textit{Proof Sketch.} We derive the bound of the expected action value (Lemma 1), the expected derivative of action value (Lemma 2), and the variance of action value (Lemma 3) over the angle-to-pole regularization. According to Assumption 1, the term $\mathbb{E}_{\tilde{\thetab}}||\tilde\phis_t-\phis^*||$ is bounded by~$\epsilon_t$. Meanwhile, by applying Lemma 3, the error term $||\sum_{j=i}^{\infty}\frac{2\beta^2\eta_j}{|\mathcal{E}_j|}\sum_{\tau\in\mathcal{E}_j}\nabla_{\phis} \expval{O}_{o, \phis, \thetab}||$ due to the regularization is upper bounded by a constant $\epsilon_t'$, completing the proof. \typeout{The detailed lemmas and proofs are deferred to Appendix.}

\subsection{Local QNN Pole Training}\label{sec:3.2} 
We design the few-shot learning in the pole domain. The reason is as follows:
First of all, if the size of QNN is small, then the QNN suffers from adapting to multiple environments. The simplest way to cope with this problem is that extend the model size. However, scaling up the QNN model size is also challenging. Second, the measurement is an excellent kernel where the entangled quantum state is mapped to the observable~\cite{havlivcek2019supervised,Schuld2022QML}, while using very small parameters. The last one is that the heritage of measurement enables fast catastrophic remembering to be more intuitive, reversible, and memorable than the classical few-shot method~\cite{farquhar2018towards}.
Motivated by the mentioned above, we propose the \textit{pole memory} where the pole parameters are temporally/permanently stored. 

For the local-QNN pole training, and inspired by \textit{VDN~\cite{SunehagLGCZJLSL18}}, we assume that the joint action-value can be expressed as the summation of local action-value across $N$ agents, which is written as  $Q_{tot}(s,a^1,a^2,\cdots,a^N)\approx \frac{1}{N}\sum_{n=1}^N\nolimits \tilde Q(o^n,a^n; \phis, \thetab^n)$, where $ \tilde Q(o^n,a^n; \phis, \thetab^n)$ denotes $n$-th agent's local \textit{Q}-network which is parametrized with the angle parameters $\phis$, and $n$-th agent's pole parameters $\thetab^n$. Note that the local-QNN pole training only focuses on training pole parameters $\Theta\triangleq \{\thetab^n\}^N_{n=1}$, and we do not consider angle noise in local-QNN pole training. We expect to maximize joint-action value by training the pole parameters (\textit{i.e.,} rotating the measurement axes).
To maximize the cumulative returns, we design a loss function for multi-agent as $\mathcal{L}_a\!(\Theta;\phis,\mathcal{E},\Theta') = \frac{1}{ |\mathcal{E}|}\sum_{\taus\sim \mathcal{E}}[r + \frac{1}{N}\cdot$ $\sum_{n=1}^N (\max\limits_{a'^n}\tilde Q(o'^{n},a'^n;\phis,\thetab'^n) -\tilde Q(o^n,a^n;\phis,\thetab^n))^2],$
% \begin{equation}
% \mathcal{L}_a\!(\Theta;\phis,\mathcal{E},\Theta') = \frac{1}{ |\mathcal{E}|}\sum_{\taus\sim \mathcal{E}}\Big[r + \frac{1}{N}\cdot
%     \sum_{n=1}^N (\max\limits_{a'^n}\tilde Q(o'^{n},a'^n;\phis,\thetab'^n) -\tilde Q(o^n,a^n;\phis,\thetab^n))^2\Big],     \label{eq:angle}
% \end{equation}
where $\taus =\expval{\mathbf{o},\mathbf{a},r,\mathbf{o}'}$ and $\Theta'\triangleq\{\thetab'^{n}\}^N_{n=1}$ stand for the transition sampled from environment and the target trainable pole parameters of whole agents, respectively.
Note that the loss gradient of $\mathcal{L}_a$ can be obtained by the parameter shift rule\typeout{, as we have mentioned in Appendix}. Since the convergence analysis of $\mathcal{L}_a$ is challenging, we show the effectiveness of the local-QNN pole training via numerical experiments.

%\subsection{Quantum Multi-Agent Meta Reinforcement Learning} 
\subsection{QM2ARL Algorithm} \label{sec:3.3}
\begin{algorithm2e}[t!]
\small
\caption{Training Procedure}\label{alg}
 Initialize parameters, $\phis\leftarrow \phis_0$, $\phis'\leftarrow \phis_0$, $\thetab \leftarrow \bm{0}$, and  $\forall\thetab^{n} \leftarrow \bm{0}$\;
% \;
 \While{Meta-QNN Angle Training}{
    Generate an episode,
$\mathcal{E} \leftarrow \left\{(\mathbf{o}_0, \mathbf{a}_0, r_1, \ldots, \mathbf{o}_{T-1}, \mathbf{a}_{T-1}, r_T) \right\}$, s.t. $a\sim\policy_{\phis,\thetab+\tilde{\thetab}}$, $\mathbf{a}\backslash a\sim (\text{random policy})$ \;
    Sampling angle noise for every step, $\tilde{\thetab} \sim \mathcal{U}[-\alpha,\alpha]$ \;
    Compute temporal difference, $\mathcal{L}(\phis;\thetab+\tilde{\thetab},\mathcal{E})$, and its gradient $\nabla_{\phis} \mathcal{L}(\phis;\thetab+\tilde{\thetab}, \mathcal{E})$\;
  Update angle parameters, $\phis \leftarrow \phis  - \eta \nabla_{\phis} \mathcal{L}(\phis;\thetab+\tilde{\thetab}, \mathcal{E})$\;
  \textbf{if} \textit{Target update period} \textbf{then} {$\phis' \leftarrow \phis$}\;
 }
 \While{Local-QNN Pole Training}{
  Generate an episode $\mathcal{E}\leftarrow\left\{(\mathbf{o}_0, \mathbf{a}_0, r_1, \ldots, \mathbf{o}_{T-1}, \mathbf{a}_{T-1}, r_T) \right\}$, s.t. $\forall a^n \sim \policy_{\phis,\thetab^n}$\;
    Compute temporal difference $\mathcal{L}_a(\Theta; \phis,\mathcal{E})$, and its gradient $\nabla_{\Theta}\mathcal{L}_a(\Theta; \phis,\mathcal{E})$\;
  Update $\Theta \leftarrow \Theta  - \eta \nabla_{\Theta} \mathcal{L}_a (\Theta;\phis,\mathcal{E})$\; 
  \textbf{if} \textit{Target update period} \textbf{then} {$\Theta' \leftarrow \Theta$}\;
 }
\end{algorithm2e}
The training algorithm is presented in Algorithm \ref{alg}. All parameters are initialized. Note that the pole parameters are initialized to $0$. Thus, the measurement axes of all qubits are the same as the $z$-axes in the initialization step. From (line 2) to (line 7), the parameters of the meta \textit{Q}-network are trained with an angle noise $\tilde{\thetab} \sim \mathcal{U}[-\alpha,+\alpha]$.
In meta-QNN angle training, only one agent (\textit{i.e.,} meta agent) is trained, where the meta agent interacts with other agents in a multi-agent environment. The action of the meta-agent is sampled from its meta policy, \textit{i.e.,} $\policy_{\phis, \thetab+\tilde{\thetab}}$, and the other agents follow other policies. The loss and its gradient are calculated and then updated. The target network is updated at a certain updating period. After the meta-QNN angle training, all pole parameters are trained to estimate the optimal joint action-value function, which is corresponding to the \textit{local QNN pole training}. In local QNN pole training, all actions are sampled with their local \textit{Q}-networks to achieve a goal of the new task. The rest of the local QNN pole training is similar to the procedure of the meta-QNN angle training except for training pole parameters with a different loss function. 

\subsection{Pole Memory for Fast Remembering Against Catastrophic Forgetting}\label{sec:3.4}
\begin{algorithm2e}[t!]
\small
    \SetCustomAlgoRuledWidth{0.44\textwidth} 
\caption{Learning Procedure for Fast Remembering}\label{alg:continual}
 \textbf{Notation.} $\thetab_p$: the pole from pole memory\;
\textbf{Initialization.} $\forall \phis,\phis'\leftarrow \phis_0$, $\forall\thetab,\thetab^{n} \leftarrow \bm{0}$\;
% \;
 \While{Meta-QNN Angle Training}{
 $\mathcal{E} \leftarrow \emptyset$\;
 \For{env $\in$ set of environment}
 {
    Generate an episode, $\mathcal{E}_{tmp}$\; 
    $\mathcal{E}\leftarrow \mathcal{E} \cup \mathcal{E}_{tmp}$\;
 }
 $\phis \leftarrow \phis  - \eta \nabla_{\phis} \mathcal{L}(\phis;\thetab+\tilde{\thetab}, \mathcal{E})$\;
  \textbf{if} \textit{Target update period} \textbf{then} {$\phis' \leftarrow \phis$}\;
 }
 \While{Training}{
 Initialize $\forall\thetab^{n} \leftarrow \thetab_p$, $\eta \leftarrow \eta_0$\;
 \textit{Local QNN Pole Training} in Algorithm 1\;
 }
\end{algorithm2e}
The pole memory in QM2ARL can be utilized to cope with catastrophic forgetting via fast remembering~\cite{PNAS}. To illustrate this, we first consider that a \textit{Q}-network is trained in one environment. After training, a \textit{Q}-network is trained in another environment, which may forget the learned experiences in the previous environment~\cite{mirzadeh2020understanding}.
To cope with such catastrophic forgetting, existing methods rely on replaying the entire previous experiences \cite{GR2022}, incurring a high computational cost. Alternatively, QM2ARL can first reload a meta-model in the pole memory, thereby fast remembering its previous environment with fewer iterations.
Precisely, as illustrated in Algorithm \ref{alg:continual}, the meta \textit{Q}-network is trained over various environments. Then, local QNN pole training is conducted for fine-tuning the meta Q-network to a specific environment. 
Finally, when re-encountering a forgotten environment during continual learning, the pole parameters can be re-initialized as the meta Q-network in the pole memory. This achieves fast remembering against catastrophic forgetting as we shall discuss in Fig. \ref{fig:continual}.

\section{Numerical Experiments}\label{sec:4}
% \subsection{Experiment Setup}\label{sec:4-1}
The numerical experiments are conducted to investigate the following four aspects.
%\BfParaa{\textit{Impact of Pole Parameters on the Meta Q-Network.}}
%\BfParaa{\textit{Impact of Angle-to-Pole Regularization.}}
%\BfParaa{\textit{Effectiveness of Pole Memory in Fast Remembering.}}
%\BfParaa{\textit{QM2ARL vs. QMARL and Classical MARL.}}  

\begin{figure}[t!]
\centering
\includegraphics[width=\columnwidth]{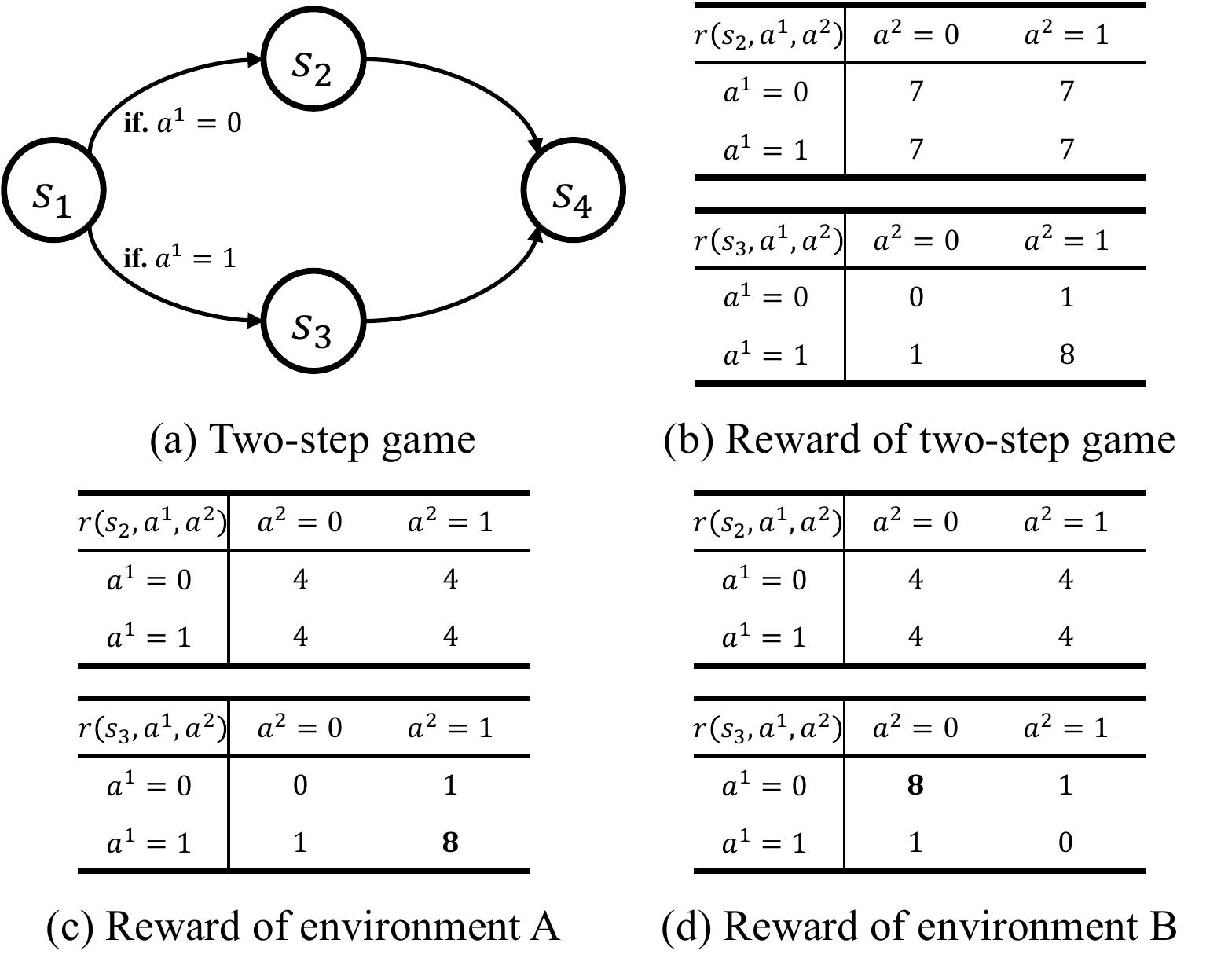}
\vspace{-20pt}
\caption{Two-step game environment.}
\label{fig:env}\vspace{-10pt}
\end{figure}
% \BfPara{Experiment Setting} For the simulation, we conduct experiments with open-source QML libraries. 
% \typeout{The simulation setting is provided in Appendix. In addition, the source codes are also submitted as supplementary material.}

\begin{figure*}[t!]
    \centering
\begin{tabular}{
                    @{}p{0.1666\linewidth}
                    @{}p{0.1666\linewidth}
                    @{}p{0.1666\linewidth}
                    @{}p{0.1666\linewidth}
                    @{}p{0.1666\linewidth}
                    @{}p{0.1666\linewidth}
                }
\multicolumn{3}{@{}c@{}}{\includegraphics[width=.5\linewidth]{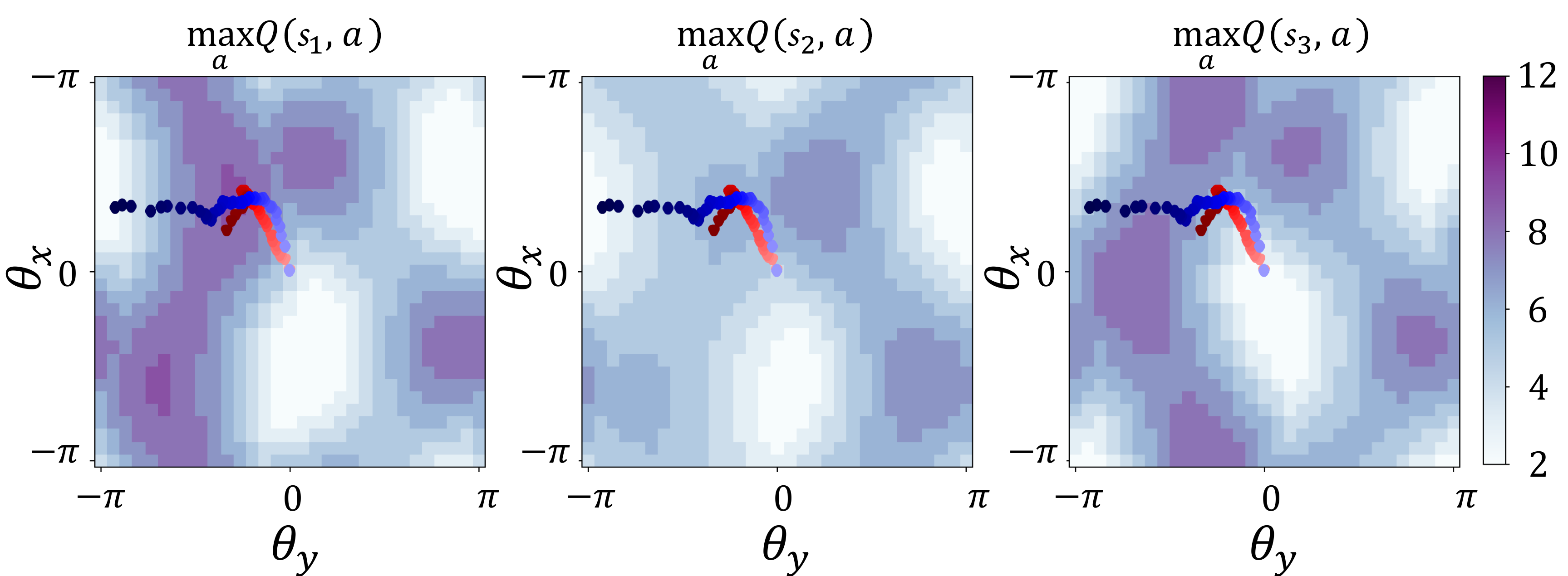}} & 
\multicolumn{3}{@{}c@{}}{\includegraphics[width=.5\linewidth]{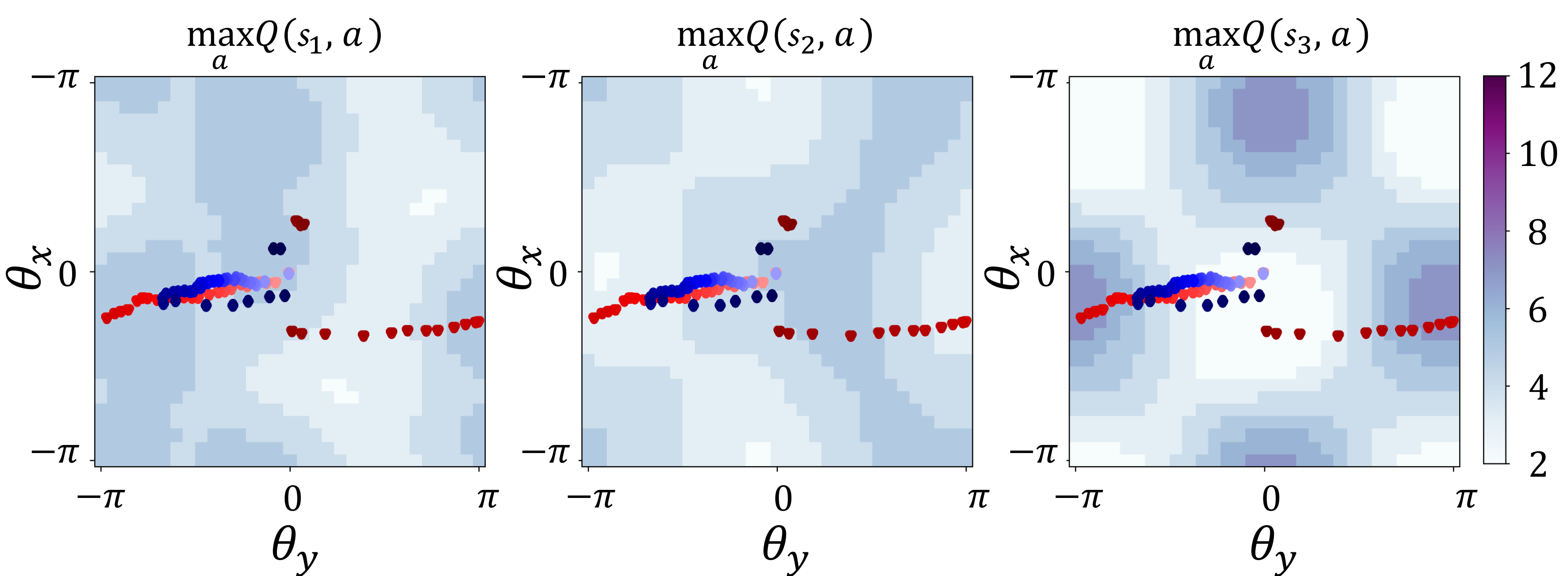}}\\
\multicolumn{3}{@{}c@{}}{(a) $\alpha=0^\circ$.} & \multicolumn{3}{@{}c@{}}{(b) $\alpha=30^\circ$.} \\ 
\multicolumn{3}{@{}c@{}}{\includegraphics[width=.5\linewidth]{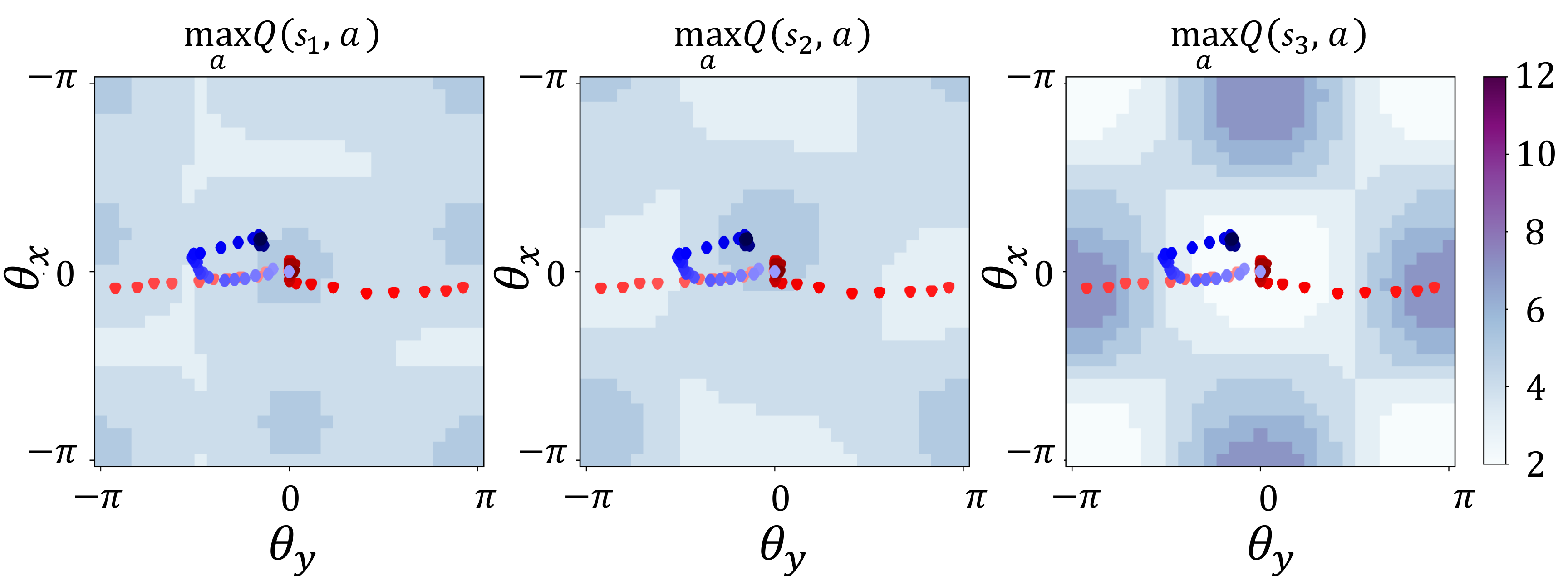}} & 
\multicolumn{3}{@{}c@{}}{\includegraphics[width=.5\linewidth]{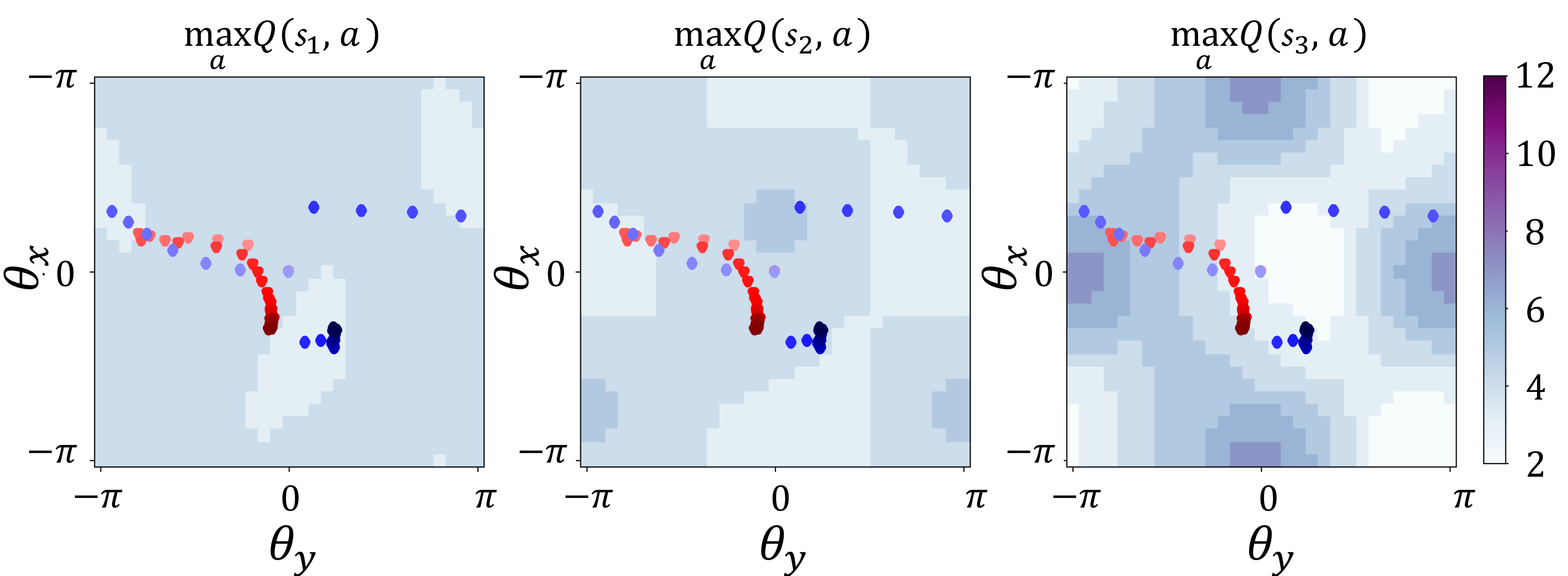}}\\
\multicolumn{3}{@{}c@{}}{(c) $\alpha=60^\circ$.} & \multicolumn{3}{@{}c@{}}{(d) $\alpha=90^\circ$.} 
\end{tabular}
\vspace{-2mm}
    \caption{Action values distribution over the pole positions in meta \textit{Q}-network and the trajectories of two agents' pole positions in the local-QNN pole training. 
    The \textit{darker} color in the grid coordinate or color bar indicates the \textit{higher} the action-value (\textit{i.e.,} $Q(s,a) = 12$), and the \textit{lighter} color is the \textit{lower} the action-value (\textit{i.e.,} $Q(s,a) = 2$). In addition, the \textit{red/blue dot} indicates the pole position of the first and second agent, respectively. In addition, the points of \textit{dark/light color} indicate the pole positions at the beginning/end of local QNN pole training.
    }
    \label{fig:qvalue}
\vspace{-3mm}
\end{figure*}

% \subsection{Numerical Results}\label{sec:4-2}
\subsection{Impact of Pole Parameters on the Meta Q-Network}
To confirm how the action-value distribution of the meta \textit{Q}-network is determined according to the position of the pole, we probe all pole positions of the meta \textit{Q}-network when the meta-QNN angle training is finished. Then, we trace the pole position while the local QNN pole training proceeds. The experiment is conducted with the two-step game environment~\cite{SonKKHY19}, which is composed of discrete state spaces and discrete action spaces\typeout{(see Appendix for their specifications)}.

Fig.~\ref{fig:qvalue} shows the action value regarding the position of the pole. Fig.~\ref{fig:qvalue}(a) corresponds to when the angle-to-pole regularization is not applied. In addition, the application of the angle-to-pole regularization is shown in Fig.~\ref{fig:qvalue}(b--d). The angle noise bound was set to $\alpha = \{ 30^\circ,60^\circ, 90^\circ\}$. 
As shown in Fig.~\ref{fig:qvalue}(a), the action-value distribution has both high and low values. When the angle-to-pole regularization exists, we figure out that the minimum and maximum values are evenly and uniformly distributed as shown in Fig.~\ref{fig:qvalue}(b--d). 
In addition, the variance of the action value is large, if angle noise exists. Therefore, the pole parameter is trained in diverse directions, and thus it can be confirmed that the momentum is large in Fig.~\ref{fig:qvalue}. 
Finally, it is obvious meta \textit{Q}-network is affected by angle-to-pole regularization.

\subsection{Impact of Angle-to-Pole Regularization}
The main proof of Theorem 1 suggests that the angle bound $\alpha$ of the angle-to-pole regularization plays a vital role in the convergence bound. 
Therefore, we conduct an experiment to investigate the role of angle-to-pole regularization. The purpose is to observe the final performance and the angle bound $\alpha$. Likewise, the experiment is conducted with a two-step game environment.

\begin{figure*}[t!]
    % \vspace{-10pt}
    \centering
    \includegraphics[width=\linewidth]{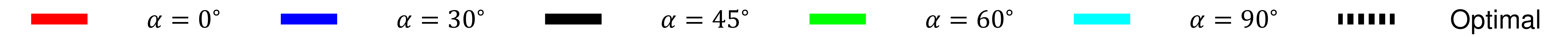}
\begin{tabular}{@{}p{0.20\linewidth}@{}p{0.20\linewidth}@{}p{0.20\linewidth}@{}p{0.20\linewidth}@{}p{0.20\linewidth}}
\includegraphics[width=\linewidth]{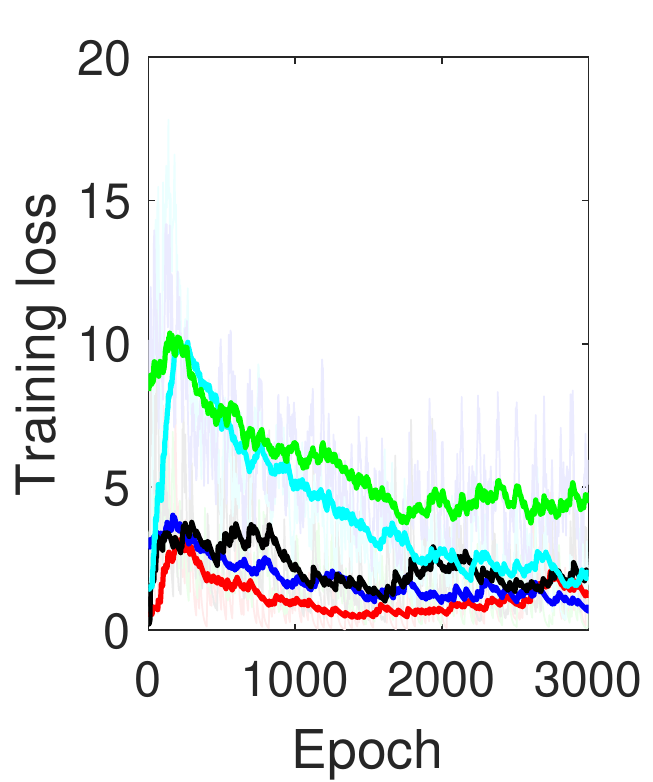}
& \includegraphics[width=\linewidth]{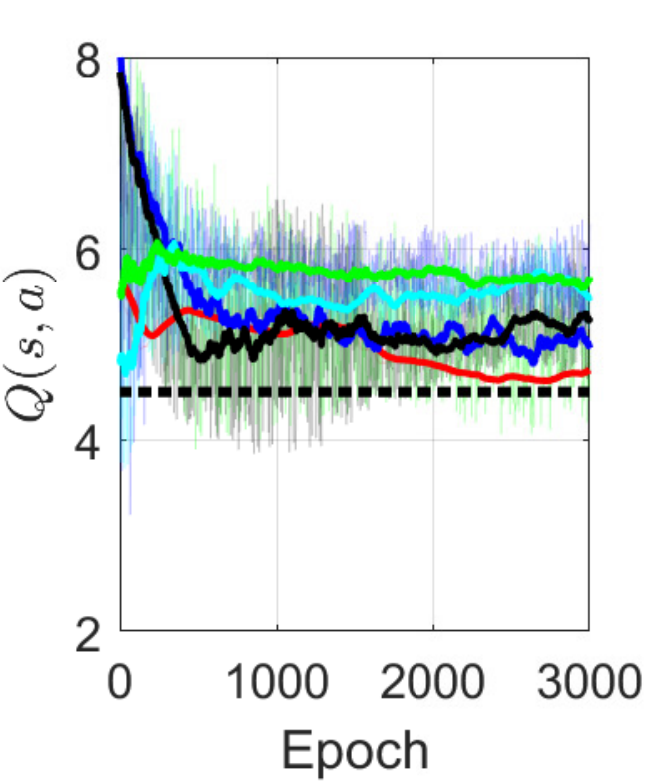}
& \includegraphics[width=\linewidth]{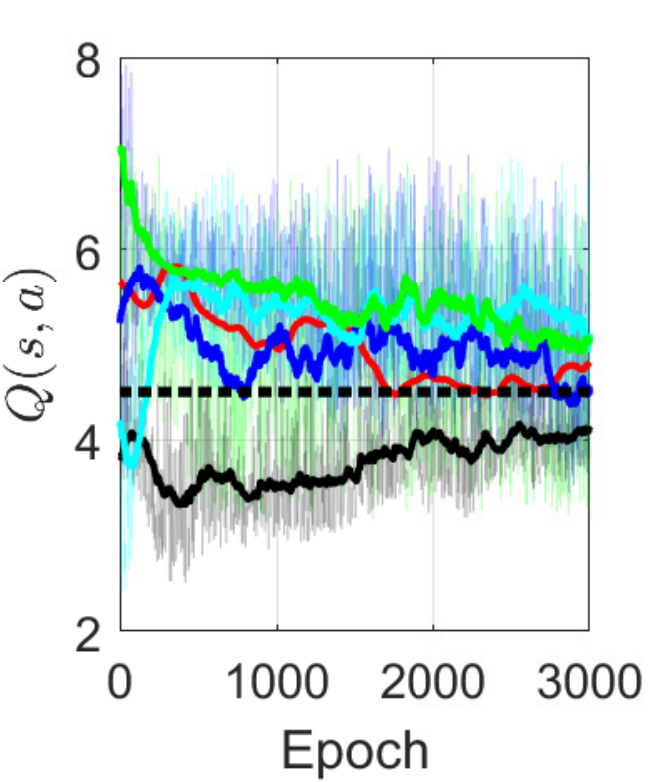}
& \includegraphics[width=\linewidth]{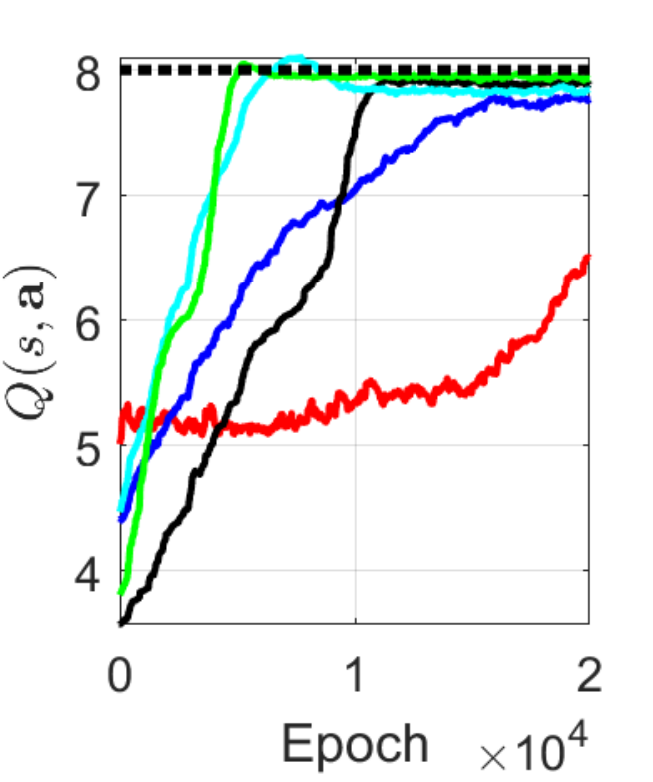}
& \includegraphics[width=\linewidth]{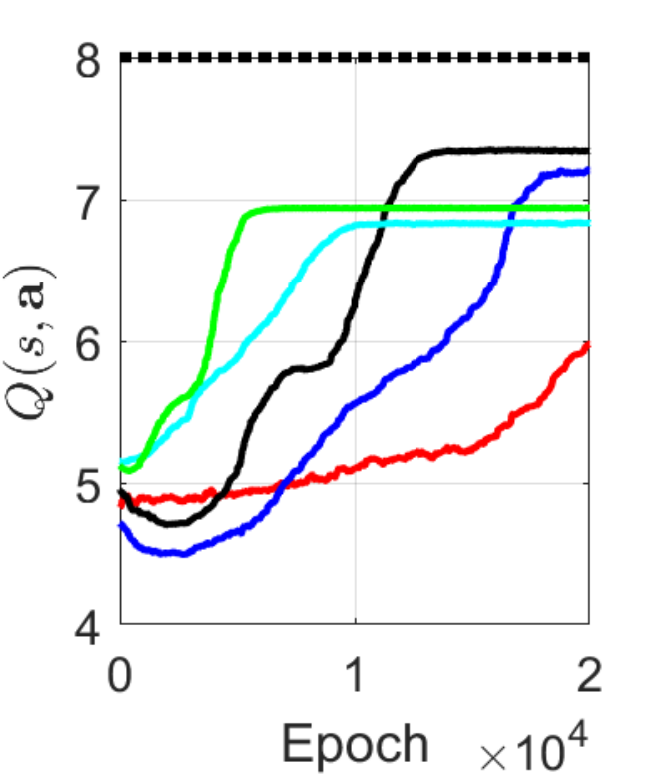}\\
  \multicolumn{1}{c}{\,} 
& \multicolumn{1}{c}{\small $Q(s_1,a=1)$.} 
& \multicolumn{1}{c}{\small $Q(s_3,a=1)$.}
& \multicolumn{1}{c}{\small $Q(s_1,\mathbf{a}=1)$.} 
& \multicolumn{1}{c}{\small $Q(s_3,\mathbf{a}=1)$.}\\
\multicolumn{1}{c}{(a) Training loss.} &
  \multicolumn{2}{c}{(b) Meta-QNN angle training.} & 
  \multicolumn{2}{c}{(c) Local QNN pole training.} 
\end{tabular}
\caption{The results of the meta-QNN angle training process in two-step game: (a) the learning curve of QM2ARL corresponding to training loss of meta-QNN angle training process, (b) the action-value of meta \textit{Q}-network given state, and (c) the joint action-value given state.}
    \label{fig:gg}
    % \vspace{-3mm}
\end{figure*}

To deep dive into the impact of the angle noise regularizer, we investigate the optimality corresponding to the loss function and optimal action-value function. We conduct meta-QNN angle training and local-QNN pole training with $3,000$ and $20,000$ iterations for the simulation, respectively. The two agents' pole parameters (\textit{i.e.,} $\thetab^1$ and $\thetab^2$) are trained in local-QNN pole training. We test under the angle noise bound $\alpha \in \{0^\circ, 30^\circ, 45^\circ, 60^\circ, 90^\circ\}$.
We set the criterion of numeric convergence when the action-values given $s_1$ and $s_3$, stop increasing/decreasing.
As shown in Fig.~\ref{fig:gg}(a), the training loss is proportioned to the intensity of angle noise. %This is because the bound of angle noise is monotonic decreasing function of $\alpha \in [0, \pi/2]$ (see Lemma 2 and its proof in Appendix \ref{app:proof_lem_2}).
%When $\alpha = 0^\circ$, there is no angle noise, thus, meta-\textit{Q} network converges. When $\alpha = 90^\circ$, angle noise interferes meta-\textit{Q} network converging.  
Fig.~\ref{fig:gg}(b)/(c) shows the numerical results of meta-QNN angle training and local QNN pole training. 
As shown in Fig.~\ref{fig:gg}(b), the larger angle bound, the distance between the action-value of the meta \textit{Q}-network and the optimal action-value is larger. 
As shown in Fig.~\ref{fig:gg}(c), the smaller angle bound, QM2ARL converges slowly, and when the angle bound is large, it shows faster convergence. 
In summary, despite the angle noise regularizer making meta-QNN angle training slow convergence, the angle noise regularizer makes the local-QNN pole training fast convergence.

\begin{figure*}[t!]
\centering
\begin{minipage}{0.33333\textwidth}
     \centering
     \begin{tabular}{@{}p{\linewidth}@{}} 
        \includegraphics[height=3.662cm]{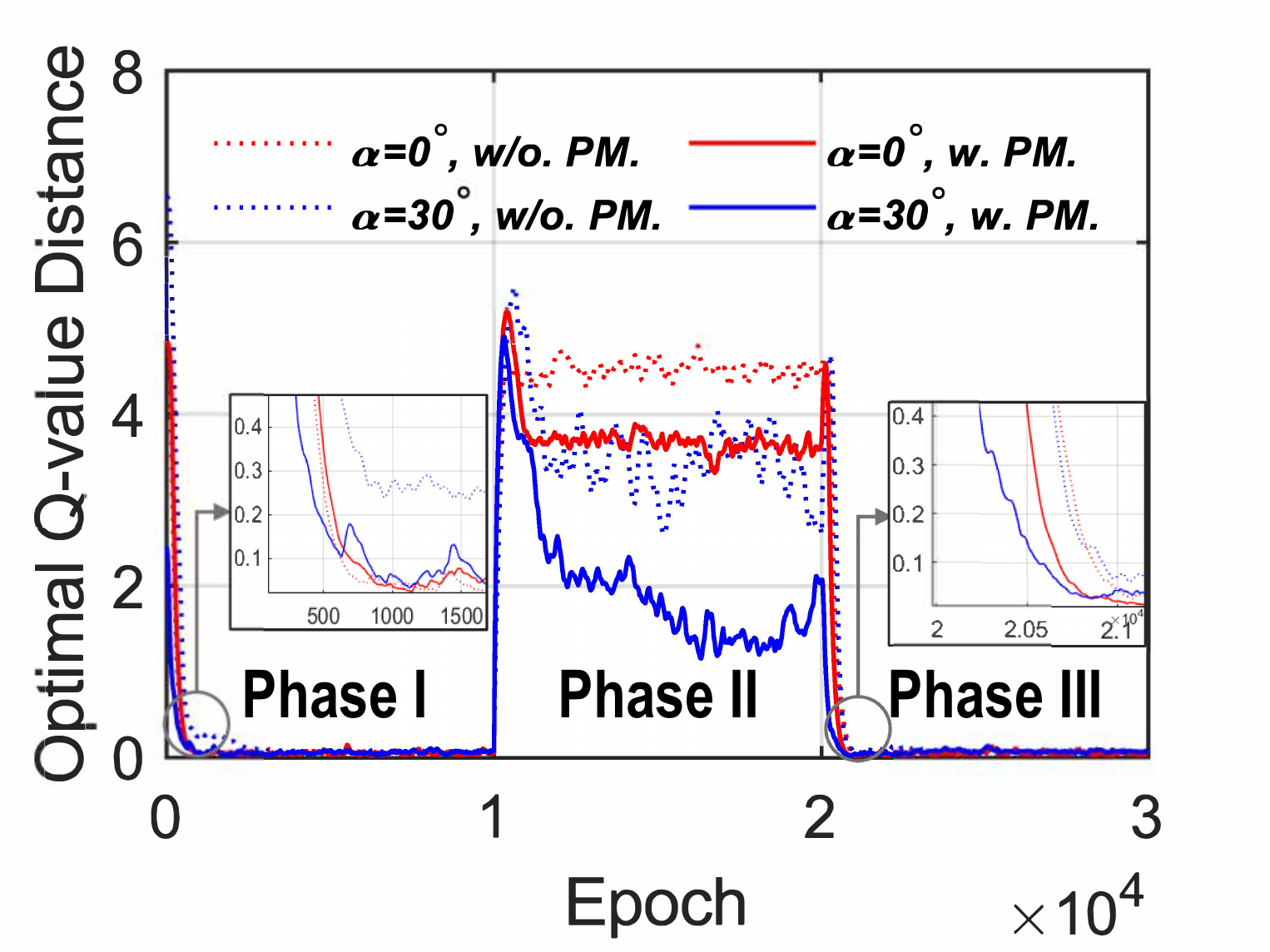}\\
        \multicolumn{1}{c}{\,} 
        \end{tabular}
        \caption{Fast remembering.}\label{fig:continual}
   \end{minipage}\hfill
   \begin{minipage}{0.66666\textwidth}
     \centering
         \begin{tabular}{@{}p{0.5\linewidth}@{}p{0.5\linewidth}@{}} 
         \includegraphics[height=3.662cm]{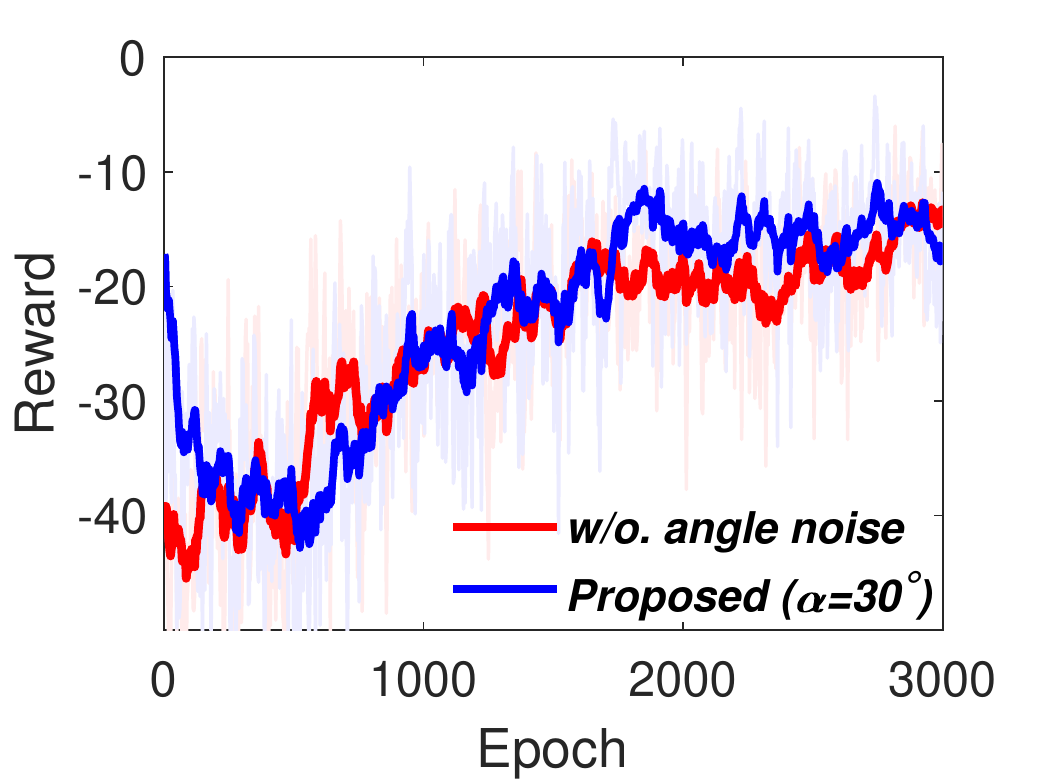} & 
         \includegraphics[height=3.662cm]{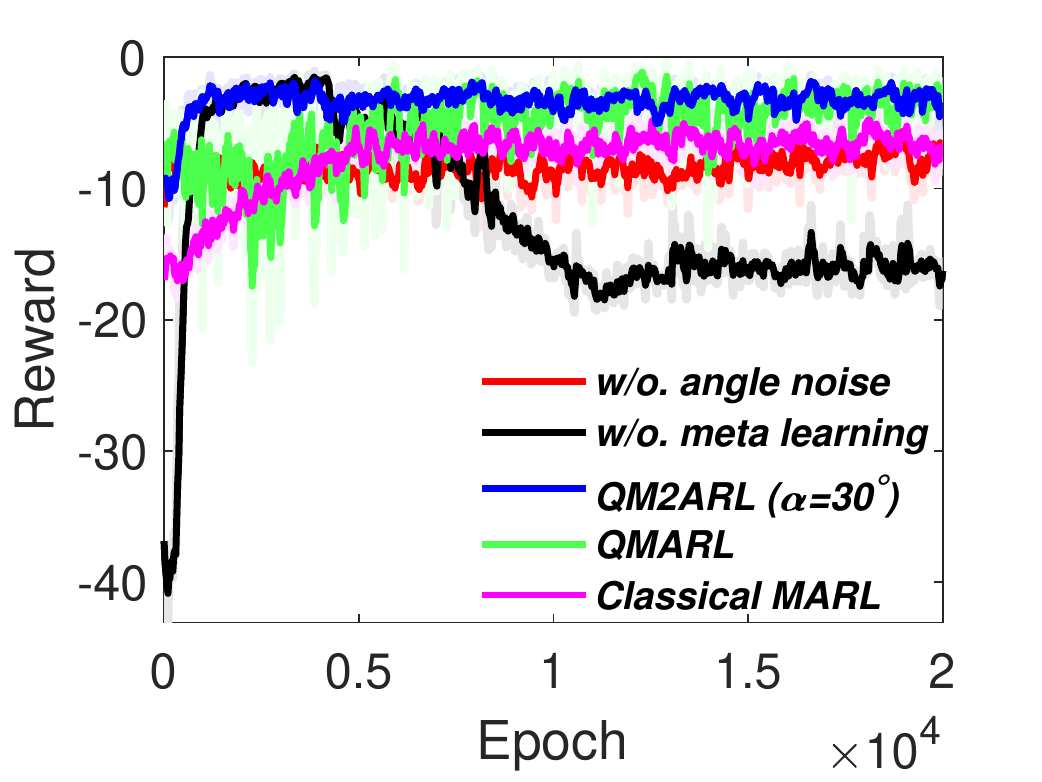}\\
        \multicolumn{1}{l}{\small \quad \text{(a)} Meta-QNN angle training.} & 
        \multicolumn{1}{l}{\small \quad \text{(b)} Local QNN pole training.} \\
    \end{tabular}   
    \caption{Performance in the \textit{single-hop offloading environment}.}
    \label{fig:singlehop}
   \end{minipage}\hfill
  \vspace{-4mm}
\end{figure*}

\subsection{Effectiveness of Pole Memory in Fast Remembering}
Compared to the Q-network used in the existing reinforcement learning, the structural advantages of QM2ARL exist corresponding to fast remembering. QM2ARL can return the pole position to the desired position. This structural property does not require a lot of memory costs, and it only requires a small number of qubits. Therefore, we investigate the advantage of QM2ARL corresponding to fast remembering. 
We consider a two-step game scenario under two different environments, Env A and Env B as shown in Fig.~\ref{fig:env}(c)/(d). To intentionally incur catastrophic forgetting, in Algorithm 2, the original reward function (i.e., Env A) is changed into a wrong reward function (Env B so that the training fails to achieve high Q-value. Then, the reward function (Env B) is rolled back to that of Env A, under which conventional training may take longer time or fail to achieve the original \textit{Q}-value. The meta \textit{Q}-network is trained for 5,000 epochs. Then, we conduct local-QNN pole training for 10,000 epochs per environment. To avoid catastrophic forgetting, we train the pole parameters in \textit{Env A} (Phase I), then \textit{Env B} (Phase II), and finally \textit{Env A} (Phase III). From this experiment, we compare QM2ARL with pole memory (denoted as w. PM.) and without pole memory (denoted as w/o. PM.) for two angles bound $\alpha  \in \{0^\circ, 30^\circ\}$. 
For performance comparison, we adopt optimal \textit{Q}-value distance\typeout{ (see Appendix). The reward function and the experimental hyper-parameters are presented in Appendix.}
Fig.~\ref{fig:continual} shows the results of the fast remembering scheme. In the dotted curve in Fig.~\ref{fig:continual}, if the value of $y$-axis is closer to zero, QM2ARL achieves adaptation to the environment. In common, all frameworks shows the better adaptation to \textit{Env A} then \textit{Env B}.
As shown in phase I of Fig.~\ref{fig:continual}, the initial tangent of the optimal \textit{Q}-value distance shows steep when the pole memory is utilized. 
In phase II, the comparison (\textit{i.e.,} no pole memory) cannot adapt to \textit{Env B}, while the proposed framework (denoted as $\alpha=30^\circ$ w. PM) adapts to \textit{Env B}. In addition, utilizing a pole memory shows faster adaptation.
The result of phase III shows faster adaptation, which is similar to the results of phase I.
In summary, the pole memory enables QM2ARL to achieve faster adaptation, and the proposed framework (\textit{i.e.,} QM2ARL leveraging both angle-to-pole regularization  and pole memory) shows faster and better adaptation.
% All comparison schemes show the similar learning tendency regarding fewshot learning in phase I, and phase III. The initial tangent of the optimal \textit{Q}-value in algorithms that leverage pole initialization, shows steeper than algorithms without pole initialization.  In phase II, there is huge performance difference corresponding to a pole initialization. The pole initialization makes QM2ARL achieve faster adaptation. Furthermore, 
% Therefore, the pole initialization has advantages in continual learning scheme.

% Then, agents' pole parameters are trained with a new task in \textit{Phase II}. In \textit{Phase III}, the angle is retrained  with the meta-task. We compare the proposed algorithm (\textit{blue dotted line} in Fig.~\ref{fig:continual}) that make the angle initialized when the task is changed. The comparisons are trained with the parameters that is trained in other tasks. The optimal value distance in Fig.~\ref{fig:continual} is calculated via \eqref{eq:distance}. As shown in Fig.~\ref{fig:continual}, the proposed method shows faster convergence in \textit{Phase III}. 

\subsection{QM2ARL vs. QMARL and Classical MARL}
We compare QM2ARL with the existing QMARL algorithms and the performance benefits compared to a classical deep \textit{Q}-network based MARL with the same number of parameters~\cite{yun2022QMARL, SunehagLGCZJLSL18}.

An experiment has been conducted on a more difficult task than a two-step game, a \textit{single-hop environment}.
In Fig.~\ref{fig:singlehop}(a), there was no significant difference in performance when angle noise exists (\textit{i.e.,} $\alpha = 30^\circ$) and when it does not exist. 
However, in the local-QNN pole training, the performance difference between the existence of angle noise is large, as shown in  \textit{red line} and \textit{blue line} of Fig.~\ref{fig:singlehop}(b).  
When comparing \textit{blue line} and \textit{green line}, the convergence of the proposed scheme is faster than that of the conventional CTDE QMARL technique. When compared with the local-QNN pole training without pretraining, the performance deterioration is significant.
In summary, like \textit{single-hop environment}, the performance of the method proposed in this paper is superior to multi-agent learning and inference over a finite horizon. 

% \balance
\section{Concluding Remarks}\label{sec:5}
Inspired from meta learning and the unique measurement nature in QML, we proposed a novel quantum MARL framework, dubbed Q2MARL. By exploiting the two trainable dimension in QML, the meta and local training processes of Q2MARL were separated in the PQC angle and its measurement pole domains, i.e.,  meta-QNN angle training followed by local-QNN pole training. By reflecting this sequential nature of angle-to-pole training operations, we developed a new angle-to-pole regularization technique that injects noise into the pole domain during angle training. Furthermore, by exploiting the angle-pole domain separation and the small pole dimension, we introduced a concept of pole memory that can save all meta-QNN and local-QNN training outcomes in the pole domain and load them only using two parameters per each. Simulation results corroborated that Q2MARL achieves higher reward with faster convergence than an QMARL baseline and a classical MARL with the same number of parameters. The results also showed that the proposed angle-to-pole regularization is effective in generalizing the meta-QNN training, yet at the cost of compromising convergence speed. It is therefore worth optimizing this trade-off between generalization and convergence speed in future research. Lastly, the simulations demonstrated the effectiveness of the proposed pole memory in enabling fast remembering against catastrophic forgetting. Investigating its more benefits in continual learning as well as incorporating multiple measurement dimensions could be interesting topics for future research.

\vspace{1mm}
\BfPara{Acknowledgments}
This work was partly supported by Institute of Information \& communications Technology Planning \& Evaluation (IITP) grant funded by the Korea government(MSIT) (No.2022-0-00907, Development of AI Bots Collaboration Platform and Self-organizing AI) and National Research Foundation of Korea (NRF-2022R1A2C2004869). Jihong Park and Joongheon Kim are the corresponding authors of this paper.
\fontsize{9.8pt}{10.8pt}\selectfont \bibliography{aaai23}

\begin{thebibliography}{32}
\providecommand{\natexlab}[1]{#1}

\bibitem[{Bloch(1946)}]{bloch1946nuclear}
Bloch, F. 1946.
\newblock Nuclear induction.
\newblock \emph{Physical review}, 70(7-8): 460.

\bibitem[{Bouwmeester and Zeilinger(2000)}]{bouwmeester2000physics}
Bouwmeester, D.; and Zeilinger, A. 2000.
\newblock The physics of quantum information: basic concepts.
\newblock In \emph{The physics of quantum information}, 1--14. Springer.

\bibitem[{Chen et~al.(2020)Chen, Yang, Qi, Chen, Ma, and Goan}]{chen20}
Chen, S. Y.-C.; Yang, C.-H.~H.; Qi, J.; Chen, P.-Y.; Ma, X.; and Goan, H.-S.
  2020.
\newblock Variational quantum circuits for deep reinforcement learning.
\newblock \emph{IEEE Access}, 8: 141007--141024.

\bibitem[{Cho(2020)}]{cho2020ibm}
Cho, A. 2020.
\newblock IBM promises 1000-qubit quantum computer—a milestone—by 2023.
\newblock \emph{Science}, 15.

\bibitem[{Daniels et~al.(2022)Daniels, Raghavan, Hostetler, Rahman, Sur,
  Piacentino, and Divakaran}]{GR2022}
Daniels, Z.; Raghavan, A.; Hostetler, J.; Rahman, A.; Sur, I.; Piacentino, M.;
  and Divakaran, A. 2022.
\newblock Model-Free Generative Replay for Lifelong Reinforcement Learning:
  Application to Starcraft-2.
\newblock In \emph{Proc. of the Conference on Lifelong Learning Agents
  (CoLLAs)}. Montreal, Canada.

\bibitem[{Farquhar and Gal(2018)}]{farquhar2018towards}
Farquhar, S.; and Gal, Y. 2018.
\newblock Towards robust evaluations of continual learning.
\newblock \emph{arXiv preprint arXiv:1805.09733}.

\bibitem[{Finn, Abbeel, and Levine(2017)}]{finn2017model}
Finn, C.; Abbeel, P.; and Levine, S. 2017.
\newblock Model-agnostic meta-learning for fast adaptation of deep networks.
\newblock In \emph{Proc. of the International Conference on Machine Learning
  (ICML)}, 1126--1135. Sydney, Austraila: PMLR.

\bibitem[{Gambetta(2022)}]{roadmap2022}
Gambetta, J. 2022.
\newblock Our new 2022 development roadmap.
\newblock \emph{IBM Quantum Computing}.

\bibitem[{Gentini et~al.(2020)Gentini, Cuccoli, Pirandola, Verrucchi, and
  Banchi}]{gentini2020noise}
Gentini, L.; Cuccoli, A.; Pirandola, S.; Verrucchi, P.; and Banchi, L. 2020.
\newblock Noise-resilient variational hybrid quantum-classical optimization.
\newblock \emph{Physical Review A}, 102(5): 052414.

\bibitem[{Harrow and Napp(2021)}]{harrow2021low}
Harrow, A.~W.; and Napp, J.~C. 2021.
\newblock Low-depth gradient measurements can improve convergence in
  variational hybrid quantum-classical algorithms.
\newblock \emph{Physical Review Letters}, 126(14): 140502.

\bibitem[{Havl{\'\i}{\v{c}}ek et~al.(2019)Havl{\'\i}{\v{c}}ek, C{\'o}rcoles,
  Temme, Harrow, Kandala, Chow, and Gambetta}]{havlivcek2019supervised}
Havl{\'\i}{\v{c}}ek, V.; C{\'o}rcoles, A.~D.; Temme, K.; Harrow, A.~W.;
  Kandala, A.; Chow, J.~M.; and Gambetta, J.~M. 2019.
\newblock Supervised learning with quantum-enhanced feature spaces.
\newblock \emph{Nature}, 567(7747): 209--212.

\bibitem[{Jerbi et~al.(2021)Jerbi, Gyurik, Marshall, Briegel, and
  Dunjko}]{jerbi2021variational}
Jerbi, S.; Gyurik, C.; Marshall, S.; Briegel, H.~J.; and Dunjko, V. 2021.
\newblock Variational quantum policies for reinforcement learning.
\newblock In \emph{Proc. of Neural Information Processing Systems (NeurIPS)}.
  Virtual.

\bibitem[{Killoran et~al.(2019)Killoran, Bromley, Arrazola, Schuld, Quesada,
  and Lloyd}]{killoran2019continuous}
Killoran, N.; Bromley, T.~R.; Arrazola, J.~M.; Schuld, M.; Quesada, N.; and
  Lloyd, S. 2019.
\newblock Continuous-variable quantum neural networks.
\newblock \emph{Physical Review Research}, 1(3): 033063.

\bibitem[{Kirkpatrick et~al.(2017)Kirkpatrick, Pascanu, Rabinowitz, Veness,
  Desjardins, Rusu, Milan, Quan, Ramalho, Grabska-Barwinska, Hassabis, Clopath,
  Kumaran, and Hadsell}]{PNAS}
Kirkpatrick, J.; Pascanu, R.; Rabinowitz, N.; Veness, J.; Desjardins, G.; Rusu,
  A.~A.; Milan, K.; Quan, J.; Ramalho, T.; Grabska-Barwinska, A.; Hassabis, D.;
  Clopath, C.; Kumaran, D.; and Hadsell, R. 2017.
\newblock Overcoming catastrophic forgetting in neural networks.
\newblock \emph{Proceedings of the National Academy of Sciences}, 114(13):
  3521--3526.

\bibitem[{Lockwood and Si(2020{\natexlab{a}})}]{lockwood2021playing}
Lockwood, O.; and Si, M. 2020{\natexlab{a}}.
\newblock Playing atari with hybrid quantum-classical reinforcement learning.
\newblock In \emph{Proc. NeurIPS Workshop on Pre-registration in Machine
  Learning}, 285--301. PMLR.

\bibitem[{Lockwood and Si(2020{\natexlab{b}})}]{lockwood2020reinforcement}
Lockwood, O.; and Si, M. 2020{\natexlab{b}}.
\newblock Reinforcement learning with quantum variational circuit.
\newblock In \emph{Proc. of the AAAI Conference on Artificial Intelligence and
  Interactive Digital Entertainment}, volume~16, 245--251.

\bibitem[{Mirzadeh et~al.(2020)Mirzadeh, Farajtabar, Pascanu, and
  Ghasemzadeh}]{mirzadeh2020understanding}
Mirzadeh, S.~I.; Farajtabar, M.; Pascanu, R.; and Ghasemzadeh, H. 2020.
\newblock Understanding the role of training regimes in continual learning.
\newblock \emph{Advances in Neural Information Processing Systems}, 33:
  7308--7320.

\bibitem[{Mitarai et~al.(2018)Mitarai, Negoro, Kitagawa, and Fujii}]{mitarai18}
Mitarai, K.; Negoro, M.; Kitagawa, M.; and Fujii, K. 2018.
\newblock Quantum circuit learning.
\newblock \emph{Physical Review A}, 98(3): 032309.

\bibitem[{Nielsen and Chuang(2002)}]{nielsen2002quantum}
Nielsen, M.~A.; and Chuang, I. 2002.
\newblock Quantum computation and quantum information.

\bibitem[{O'Brien et~al.(2004)O'Brien, Pryde, Gilchrist, James, Langford,
  Ralph, and White}]{o2004quantum}
O'Brien, J.~L.; Pryde, G.; Gilchrist, A.; James, D.; Langford, N.~K.; Ralph,
  T.; and White, A. 2004.
\newblock Quantum process tomography of a controlled-NOT gate.
\newblock \emph{Physical review letters}, 93(8): 080502.

\bibitem[{Oliehoek and Amato(2016)}]{Springer2016_POMDP}
Oliehoek, F.~A.; and Amato, C. 2016.
\newblock \emph{A Concise Introduction to Decentralized {POMDPs}}.
\newblock Springer Publishing Company, Incorporated.

\bibitem[{Rashid et~al.(2020)Rashid, Samvelyan, de~Witt, Farquhar, Foerster,
  and Whiteson}]{RashidSWFFW20}
Rashid, T.; Samvelyan, M.; de~Witt, C.~S.; Farquhar, G.; Foerster, J.~N.; and
  Whiteson, S. 2020.
\newblock Monotonic Value Function Factorisation for Deep Multi-Agent
  Reinforcement Learning.
\newblock \emph{Journal of Machine Learning Research}, 21: 178:1--178:51.

\bibitem[{Schuld et~al.(2019)Schuld, Bergholm, Gogolin, Izaac, and
  Killoran}]{schuld19}
Schuld, M.; Bergholm, V.; Gogolin, C.; Izaac, J.; and Killoran, N. 2019.
\newblock Evaluating analytic gradients on quantum hardware.
\newblock \emph{Physical Review A}, 99(3): 032331.

\bibitem[{Schuld et~al.(2020)Schuld, Bocharov, Svore, and
  Wiebe}]{schuld2020circuit}
Schuld, M.; Bocharov, A.; Svore, K.~M.; and Wiebe, N. 2020.
\newblock Circuit-centric quantum classifiers.
\newblock \emph{Physical Review A}, 101(3): 032308.

\bibitem[{Schuld and Killoran(2019)}]{schuld2019quantum}
Schuld, M.; and Killoran, N. 2019.
\newblock Quantum machine learning in feature Hilbert spaces.
\newblock \emph{Physical review letters}, 122(4): 040504.

\bibitem[{Schuld and Killoran(2022)}]{Schuld2022QML}
Schuld, M.; and Killoran, N. 2022.
\newblock Is Quantum Advantage the Right Goal for Quantum Machine Learning?
\newblock \emph{PRX Quantum}, 3: 030101.

\bibitem[{Son et~al.(2019)Son, Kim, Kang, Hostallero, and Yi}]{SonKKHY19}
Son, K.; Kim, D.; Kang, W.~J.; Hostallero, D.; and Yi, Y. 2019.
\newblock {QTRAN}: Learning to Factorize with Transformation for Cooperative
  Multi-Agent Reinforcement Learning.
\newblock In Chaudhuri, K.; and Salakhutdinov, R., eds., \emph{Proc. of the
  International Conference on Machine Learning ({ICML})}, volume~97,
  5887--5896. Long Beach, CA, USA.

\bibitem[{Sunehag et~al.(2018)Sunehag, Lever, Gruslys, Czarnecki, Zambaldi,
  Jaderberg, Lanctot, Sonnerat, Leibo, Tuyls, and Graepel}]{SunehagLGCZJLSL18}
Sunehag, P.; Lever, G.; Gruslys, A.; Czarnecki, W.~M.; Zambaldi, V.~F.;
  Jaderberg, M.; Lanctot, M.; Sonnerat, N.; Leibo, J.~Z.; Tuyls, K.; and
  Graepel, T. 2018.
\newblock Value-Decomposition Networks For Cooperative Multi-Agent Learning
  Based On Team Reward.
\newblock In \emph{Proc. of the International Conference on Autonomous Agents
  and MultiAgent Systems, ({AAMAS})}, 2085--2087. Stockholm, Sweden.

\bibitem[{Tampuu et~al.(2017)Tampuu, Matiisen, Kodelja, Kuzovkin, Korjus, Aru,
  Aru, and Vicente}]{tampuu2017multiagent}
Tampuu, A.; Matiisen, T.; Kodelja, D.; Kuzovkin, I.; Korjus, K.; Aru, J.; Aru,
  J.; and Vicente, R. 2017.
\newblock Multiagent cooperation and competition with deep reinforcement
  learning.
\newblock \emph{PloS one}, 12(4): e0172395.

\bibitem[{Van~Hasselt, Guez, and Silver(2016)}]{van2016deep}
Van~Hasselt, H.; Guez, A.; and Silver, D. 2016.
\newblock Deep reinforcement learning with double q-learning.
\newblock In \emph{Proc. of the AAAI Conference on Artificial Intelligence},
  volume~30. Arizona, USA.

\bibitem[{Wilson et~al.(2021)Wilson, Stromswold, Wudarski, Hadfield, Tubman,
  and Rieffel}]{wilson2021optimizing}
Wilson, M.; Stromswold, R.; Wudarski, F.; Hadfield, S.; Tubman, N.~M.; and
  Rieffel, E.~G. 2021.
\newblock Optimizing quantum heuristics with meta-learning.
\newblock \emph{Quantum Machine Intelligence}, 3(1): 1--14.

\bibitem[{Yun et~al.(2022)Yun, Kwak, Kim, Cho, Jung, Park, and
  Kim}]{yun2022QMARL}
Yun, W.~J.; Kwak, Y.; Kim, J.~P.; Cho, H.; Jung, S.; Park, J.; and Kim, J.
  2022.
\newblock Quantum Multi-Agent Reinforcement Learning via Variational Quantum
  Circuit Design.
\newblock In \emph{Proc. of the IEEE International Conference on Distributed
  Computing Systems (ICDCS)}. Bologna, Italy.

\end{thebibliography}
%\counterwithin{equation}{section}
%\counterwithin{figure}{section}
%\counterwithin{table}{section}
\setcounter{page}{1}
\setcounter{table}{0}    
\setcounter{figure}{0}   
\newpage
\onecolumn
\vbox{%
    \hsize\textwidth
    \linewidth\hsize
    \hrule height 4pt
    \vskip 0.25in
    \vskip -\parskip%
    \centering
    {\LARGE\bf Supplementary Materials for:\\ {\fontsize{15}{15} \selectfont Quantum Multi-Agent Meta Reinforcement Learning}\par} 
    \vskip 0.29in
    \vskip -\parskip
    \hrule height 1pt
    \vskip 0.3in%
  }
  
\vbox{
    \centering
    {\bf{\fontsize{11}{11} \selectfont Won Joon Yun, Jihong Park, Joongheon Kim}\par}
    \vskip 0.1in
}
% \section*{Appendices}
\paragraph{Outline}
The supplementary material is organized as follows.
We describe basic quantum computing and quantum gates, we present the derivation of loss gradient using parameter shift rule, and we provide the derivation of Theorem 1. We show detailed simulation settings. Then, we describe the two-step game where the experiments are conducted. 
We provide additional information on continual learning. Next, we provide information on the single-hop offloading environment for the generalizability of QM2ARL. Finally, we provide additional experiment results.

\section*{Basic Quantum Computing and Quantum Gates}\label{app:basicquantum}
\begin{wrapfigure}[15]{r}{0.3\textwidth}
    \centering         \includegraphics[width=.9\linewidth]{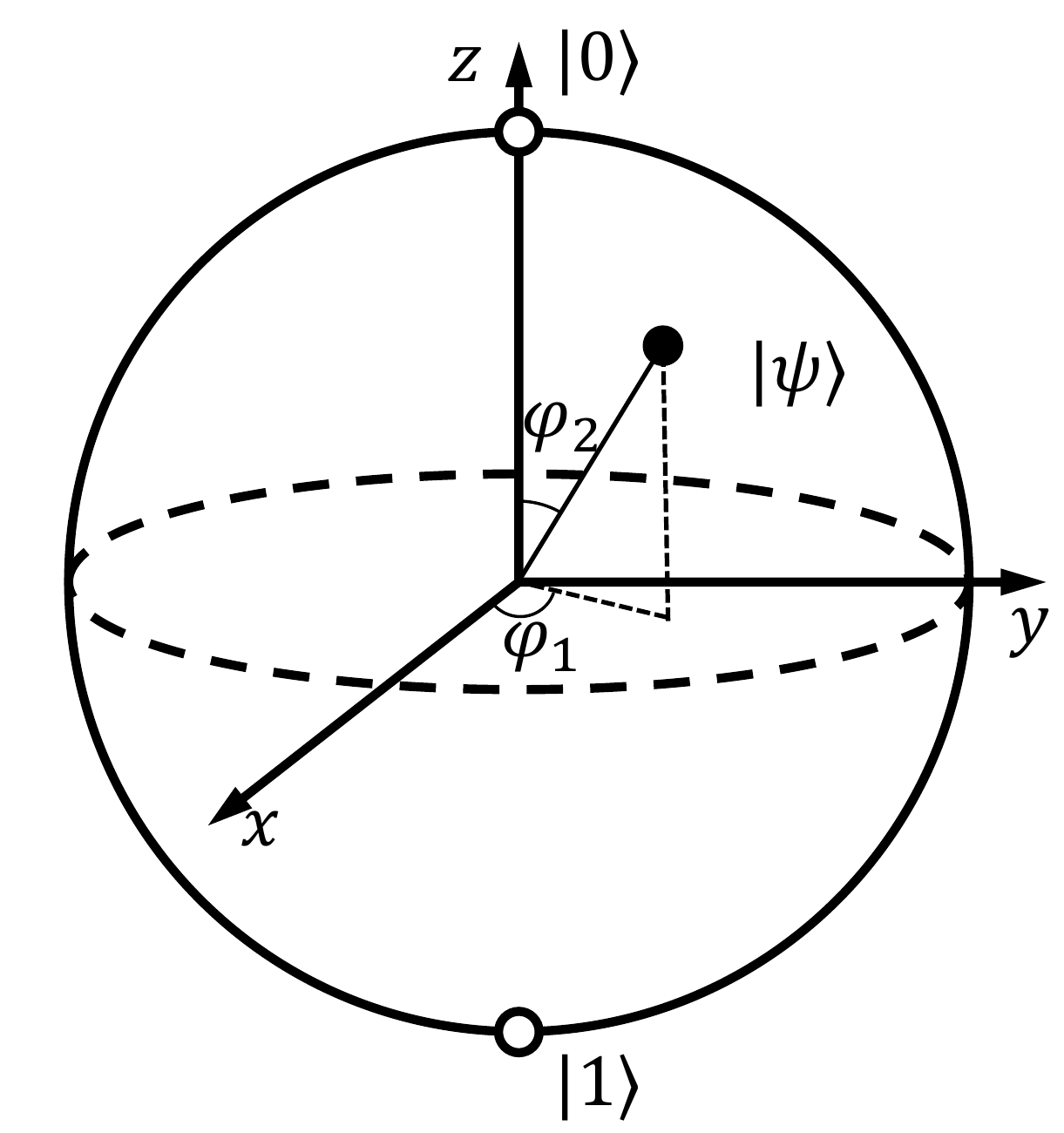}
    \caption{Bloch sphere.}
    \label{fig:bloch_representation}
\end{wrapfigure}
In quantum physics, quantum mechanics is mathematically formulated in Hilbert space, and it can be represented as \textit{Bloch sphere}. The quantum state is expressed by two standard basis vectors which are denoted as $|0\rangle$ and $|1\rangle$. A qubit is a basic unit of quantum information~\cite{bouwmeester2000physics}, and the quantum state is written as $\ket{\psi} = \cos \left(\varphi_1 /2\right)|0\rangle \,+\,e^{i\varphi_2 }\sin \left(\varphi_1 /2\right)|1\rangle$, $\forall \varphi_1, \varphi_2 \in \mathbb{R}[-\pi,\pi]$. 
A quantum gate is a unitary operation $U$ acting on Hilbert space $\mathcal{H}$. When a gate $U$ acts non-trivially only on a subset $S \subseteq [L]$ of qubits, we identify it to the operation $U\otimes \mathbbm{1}_{[L]\backslash S}$. The single-qubit Pauli gates $X$, $Y$, and $Z$ are defined as follows,
\begin{equation}
X 
= \begin{bmatrix}
0 & 1 \\ 1 & 0
\end{bmatrix},\quad Y 
= \begin{bmatrix} 0 & -i \\ i & 0 \end{bmatrix},\quad  Z = 
\begin{bmatrix} 1 & 0 \\ 0 & -1 \end{bmatrix},
\end{equation}
 and their associated rotations $R_x$, $R_y$, and $R_z$ are defined as $
 R_x(\delta) = \exp(-i \frac{\delta}{2} X), \quad R_y(\delta) = \exp(-i \frac{\delta}{2} Y), \quad R_z(\delta) = \exp(-i \frac{\delta}{2} Z),$
where rotation angles are denoted as $\delta\in\mathbb{R}[-\pi,\pi]$. A QNN utilizes these gates and controlled-NOT gates for neural network modeling and computation.

\section{Derivation of Loss Gradient}\label{app:parametershift}
The loss function is calculated with the temporal difference of meta \textit{Q}-network, which is given by,
$\mathcal{L}(\phis;\thetab, \mathcal{E}) = \frac{1}{|\mathcal{E}|}\sum_{\expval{o,a,r,o'} \in \mathcal{E}}(r+\max_{a'} Q(o',a';\phis',\thetab) -  Q(o,a;\phis,\thetab))^2$,
where $o'$, $a'$, $\phis'$, $\mathcal{E}$ and $|\mathcal{E}|$ denote the next observation, next action, target angle parameters, transitions set composing the episode and the number of transitions in the episode.
The gradient of the observable is obtained by parameter shift rule~\cite{mitarai18,schuld19}, which is written as 
\begin{equation}\label{eq:paramshift}
    \frac{\partial \expval{O_a}_{o,\phis,\thetab} }{\partial{\phi_k}}  = \frac{\expval{O_a}_{o,\phis + c_k\mathbf{e}_k ,\thetab} -\expval{O_a}_{o,\phis - c_k\mathbf{e}_k,\thetab}}{2|c_k|},
\end{equation} where $c_k\in \mathbb{R}$ and $\mathbf{e}_k$ denote a small number and $k$-th standard basis.
The initial form of loss gradient is written as follows:
\begin{equation}
    \frac{\partial \mathcal{L}(\phis;\thetab, \mathcal{E})}{\partial\phi_k} = - \frac{2}{|\mathcal{E}|}\sum_{\expval{o,a,r,o'} \in \mathcal{E}}\Big(r+\max_{a'} Q(o',a';\phis',\thetab) - Q(o,a;\phis,\thetab)\Big) \cdot  \frac{\partial Q(o,a;\phis,\thetab)}{\partial \phi_k}
\end{equation}
Substituting $\frac{\partial \expval{O_a}_{o,\phis,\thetab} }{\partial{\phi_k}}$ into $\frac{\partial \mathcal{L}(\phis;\thetab) }{\partial{\phi_k}} $, the gradient of angle parameter $\phi_k$ is derived as follows:
\begin{align}
\nonumber
    \frac{\partial \mathcal{L}(\phis;\thetab, \mathcal{E})}{\partial\phi_k}=&-\frac{2\beta^2}{|\mathcal{E}|}\sum_{\expval{o,a,r,o'} \in \mathcal{E}} \Big(\frac{r}{\beta}+\max_{a'}\expval{O_{a'}}_{o', \phis', \thetab} -\frac{\expval{O_a}_{o, \phis + c_k\mathbf{e}_k, \thetab} + \expval{O_a}_{o, \phis - c_k\mathbf{e}_k, \thetab}}{2}\Big)\\\label{eq:app-B}
    & \cdot \frac{\partial \expval{O_a}_{o,\phis,\thetab} }{\partial{\phi_k}}.
\end{align}
Since $\frac{\expval{O_a}_{o, \phis + c_k\mathbf{e}_k, \thetab} + \expval{O_a}_{o, \phis - c_k\mathbf{e}_k, \thetab}}{2} \approx \expval{O_a}_{o,\phis,\thetab}$, \eqref{eq:app-B} can be written as follows:
\begin{equation}\label{eq:lossgradient_entry}
    \frac{\partial \mathcal{L}(\phis;\thetab, \mathcal{E})}{\partial\phi_k}=-\frac{2\beta^2}{|\mathcal{E}|}\sum_{\expval{o,a,r,o'} \in \mathcal{E}} \Big(\frac{r}{\beta}+\max_{a'}\expval{O_{a'}}_{o', \phis', \thetab} - \expval{O_a}_{o,\phis,\thetab} \Big) \cdot \frac{\partial \expval{O_a}_{o,\phis,\thetab} }{\partial{\phi_k}}.
\end{equation}
It can be written as the vectorized form of partial derivatives, \textit{i.e.,} \begin{equation}
\nabla_{\phis}\mathcal{L}(\phis;\thetab,\mathcal{E}) \triangleq \big[
     \cdots\frac{\partial \mathcal{L}(\phis;\thetab,\mathcal{E}) }{\partial{\phi_k}}\cdots\big]^T.
\end{equation}
     
Then we finalize the loss gradient as follows: 
 \begin{eqnarray}\label{eq:lossgradient_final}
    \nabla_{\phis}\mathcal{L}(\phis;\thetab, \mathcal{E}) &=&-\frac{2\beta^2}{|\mathcal{E}|}\sum\limits_{\underbrace{\tiny \expval{o,a,r,o'}}_{\tau} \in \mathcal{E}} \underbrace{\Big(\frac{r}{\beta}+\max_{a'}\expval{O_{a'}}_{o', \phis', \thetab} - \expval{O_a}_{o,\phis,\thetab} \Big)}_{=A_1(\tau)} \cdot \nabla_{\phis}\expval{O_a}_{o,\phis,\thetab}, \nonumber \\
    &=&-\frac{2\beta^2}{|\mathcal{E}|}\sum_{\tau \in \mathcal{E}}A_1(\tau)\cdot \nabla_{\phis}\expval{O_a}_{o,\phis,\thetab}.
\end{eqnarray}
\section{Proof of Theorem 1}\label{app:theorem}
The lemmas below strongly imply the convergence of meta \textit{Q}-network with angle noise regularizer when the original one converges. This idea will be detailed in Theorem 1.

\subsection{Lemma 1: \textit{Expectation of meta Q-network with angle noise regularizer}}\label{app:proof_lem_1}
\begin{lemma}\label{lem:lemma1} The expectation of meta \textit{Q}-network is derived as the multiplication form of function of $\alpha$ and meta \textit{Q}-network, i.e., 
	$\mathbb{E}_{\tilde{\thetab} \sim \mathcal{U}}[Q(o,a;\phis,\thetab+\tilde{\thetab})] =\frac{\sin\alpha}{\alpha}Q(o,a;\phis,\thetab)$.
\end{lemma} 
\begin{proof}
The expectation of the difference between action value function with angle noise  and without noise is written as, 
\begin{equation}\label{eq:1}
    \mathbb{E}_{\tilde{\thetab} \sim \mathcal{U}}[Q(o,a;\phis,\thetab+\tilde{\thetab})-Q(o,a;\phis,\thetab)]=\beta\ev{\mathbb{E}_{\tilde{\thetab} \sim \mathcal{U}}[\sum_{m \in \mathcal{M}_a} P_{m,\thetab+\tilde \thetab}-P_{m,\thetab}]}{\psi_{o,\phis}}
\end{equation}
The expectation of the difference between projection with an angle noise and without noise can be written as,
\begin{align}\label{eq:2}
    \mathbb{E}_{\tilde{\theta}_m}[M_{\theta_m + \tilde \theta_m }-M_{\theta_m}]&= 
    \mathbb{E}_{\tilde \theta_m }[R^\dagger_{y}(\theta_m) R^\dagger_{y}(\tilde \theta_m) Z R_{y}(\tilde \theta_m) R_{y}(\theta_m)- R^\dagger_{y}(\theta_m) Z R_{y}(\theta_m)],\nonumber \\
    &=R^\dagger_{y}(\theta_m)\mathbb{E}_{\tilde \theta_m}[R^\dagger_{y}(\tilde \theta_m) Z R_{y}(\tilde \theta_m)]  R_{y}(\theta_m)- R^\dagger_{y}(\theta_m) Z R_{y}(\theta_m).
\end{align}

The expectation of the projection matrix with $\tilde{\theta}$, i.e., $\mathbb{E}_{\tilde{\theta}_i}[R^\dagger_{y}(\tilde{\theta}_i) Z R_{y}(\tilde{\theta}_i)]$, can be derived as,
\begin{eqnarray}
\frac{1}{2\alpha}\int^\alpha_{-\alpha}\begin{bmatrix}
    \cos\tilde{\theta}_m & - \sin\tilde{\theta}_m \\
    -\sin\tilde{\theta}_m & \cos\tilde{\theta}_m 
    \end{bmatrix}\cdot  Z
    \text{d}\tilde{\theta}_m = \frac{\sin\alpha}{\alpha} \cdot Z.
\end{eqnarray}
Then, we derive the LHS of \eqref{eq:2} in a closed-form as follows,
\begin{equation}\label{eq:7}
    \mathbb{E}_{\tilde{\theta}_m}[{M}_{\theta_m+\tilde{\theta}_m}-{M}_{\theta_m}] = \left(\frac{\sin\alpha}{\alpha}-1\right) R^\dagger_{y}({\theta}_m) Z R_{y}({\theta}_m).
\end{equation} 
Substitute \eqref{eq:7} to \eqref{eq:1}, we finalize the proof.
\end{proof}
\subsection{Lemma 2: \textit{Expectation of meta Q-network's gradient with angle noise regularizer}}\label{app:proof_lem_2}
\begin{lemma}\label{lem:lemma2}
The expectation of parameters of meta \textit{Q}-network's gradient with angle noise is written as \begin{equation}\label{eq:lem2}
\mathbb{E}_{\tilde{\theta}}[\frac{\partial \mathcal{L}(\phis;\thetab+\tilde{\thetab},\mathcal{E}) }{\partial{\phi_k}}] \leq -\frac{2\beta^2}{|\mathcal{E}|}\frac{\sin\alpha}{\alpha}\sum_{\tau \in \mathcal{E}} A_2(\tau) \frac{\partial \expval{O_a}_{o,\phis,\thetab} }{\partial{\phi_k}},
\end{equation}
where $A_2(\tau)$ is written as,
\begin{equation}
    A_2(\tau)=\begin{cases}
    \frac{r}{\beta}+\max_{a'}\expval{O_{a'}}_{o', \phis', \thetab} -1, & \textrm{if.}~~\frac{\partial \expval{O_a}_{o,\phis,\thetab+\tilde \thetab} }{\partial{\phi_k}} \geq 0  \\
   \frac{r}{\beta}+\max_{a'}\expval{O_{a'}}_{o', \phis', \thetab} +1, & \textrm{otherwise.} 
    \end{cases}.
\end{equation}
\end{lemma}
\begin{proof}
For the simplicity, we denote $\tau = \langle o, a, r, o'\rangle$.
\eqref{eq:lossgradient_entry} with angle noise is written as follows:
\begin{eqnarray}
    \mathbb{E}_{\tilde{\theta}}[\frac{\partial \mathcal{L}(\phis;\thetab+\tilde{\thetab},\mathcal{E}) }{\partial{\phi_k}}] 
    &=& -\frac{2\beta^2}{|\mathcal{E}|}\sum_{\tau \in \mathcal{E}}\mathbb{E}_{\tilde{\theta}}\Big[ \Big(\frac{r}{\beta}+\max_{a'}\expval{O_{a'}}_{o', \phis', \thetab} - \expval{O_a}_{o,\phis,\thetab+\tilde \thetab} \Big) \cdot \frac{\partial \expval{O_a}_{o,\phis,\thetab+\tilde \thetab} }{\partial{\phi_k}}\Big]\nonumber \\
   &=& - \frac{2\beta^2}{|\mathcal{E}|}\sum_{\tau \in \mathcal{E}}
    \mathbb{E}_{\tilde{\theta}}\Big[\underbrace{\Big(\frac{r}{\beta}+\max_{a'}\expval{O_{a'}}_{o', \phis', \thetab}\Big)\cdot\frac{\partial \expval{O_a}_{o,\phis,\thetab+\tilde \thetab} }{\partial{\phi_k}}}_{\circled{1}}\Big] + \nonumber \\
    &&
    \frac{2\beta^2}{|\mathcal{E}|}\sum_{\tau \in \mathcal{E}}\mathbb{E}_{\tilde{\theta}}\Big[\underbrace{\expval{O_a}_{o,\phis,\thetab+\tilde \thetab} \cdot \frac{\partial \expval{O_a}_{o,\phis,\thetab+\tilde \thetab} }{\partial{\phi_k}}}_{\circled{2}} \Big]\label{eq:lem2_12}
\end{eqnarray}
According to Lemma~\ref{lem:lemma1} and \eqref{eq:paramshift}, the expectation of meta \textit{Q}-network's gradient is written as follows:
\begin{equation}
    \circled{1} = \frac{\sin\alpha}{\alpha}\cdot\Big(\frac{r}{\beta}+\max_{a'}\expval{O_{a'}}_{o', \phis', \thetab} - \expval{O_a}_{o,\phis,\thetab+\tilde \thetab} \Big)\cdot\frac{\partial \expval{O_a}_{o,\phis,\thetab} }{\partial{\phi_k}}.
\end{equation} 
Since, $\expval{O_a}_{o,\phis,\thetab+\tilde \thetab} \leq 1$, $\mathbb{E}_{\tilde{\theta}}[\circled{2}] \leq \mathbb{E}_{\tilde{\theta}}[|\frac{\partial \expval{O_a}_{o,\phis,\thetab+\tilde \thetab} }{\partial{\phi_k}}|] = |\mathbb{E}_{\tilde{\theta}}[\frac{\partial \expval{O_a}_{o,\phis,\thetab+\tilde \thetab} }{\partial{\phi_k}}]|$. 
Therefore, \eqref{eq:lem2_12} can be rewritten as follows:
\begin{eqnarray}
\mathbb{E}_{\tilde{\theta}}[\frac{\partial \mathcal{L}(\phis;\thetab+\tilde{\thetab},\mathcal{E}) }{\partial{\phi_k}}]&\leq&
- \frac{2\beta^2}{|\mathcal{E}|}\sum_{\tau \in \mathcal{E}}\mathbb{E}_{\tilde{\theta}}\Big[{\circled{1}}\Big] +\frac{2\beta^2}{|\mathcal{E}|}\sum_{\tau \in \mathcal{E}} \Big|\mathbb{E}_{\tilde{\theta}}\Big[\frac{\partial \expval{O_a}_{o,\phis,\thetab+\tilde \thetab} }{\partial{\phi_k}}\Big]\Big|,\nonumber \\
&=&-\frac{2\beta^2}{|\mathcal{E}|}\frac{\sin\alpha}{\alpha}\sum_{\tau \in \mathcal{E}} A_2(\tau) \frac{\partial \expval{O_a}_{o,\phis,\thetab} }{\partial{\phi_k}},
\end{eqnarray}
where $A_2(\tau)$ is written as,
\begin{equation}
    A_2(\tau)=\begin{cases}
    \frac{r}{\beta}+\max_{a'}\expval{O_{a'}}_{o', \phis', \thetab} -1, & \text{if.} \frac{\partial \expval{O_a}_{o,\phis,\thetab+\tilde \thetab} }{\partial{\phi_k}} \geq 0,  \\
   \frac{r}{\beta}+\max_{a'}\expval{O_{a'}}_{o', \phis', \thetab} +1, & \text{otherwise.} 
    \end{cases}
\end{equation}
% Then we approximate $\circled{2}$ as follows: 
% \begin{align}
% &\mathbb{E}_{\tilde{\theta}}\Big[\expval{O_a}_{o,\phis,\thetab+\tilde \thetab} \cdot \frac{\partial \expval{O_a}_{o,\phis,\thetab+\tilde \thetab} }{\partial{\phi_k}} \Big]
% \approx 
% \mathbb{E}_{\tilde{\theta}}\Big[\frac{\expval{O_a}_{o, \phis + c_k\mathbf{e}_k, \thetab} + \expval{O_a}_{o, \phis - c_k\mathbf{e}_k, \thetab}}{2}\cdot \frac{\partial \expval{O_a}_{o,\phis,\thetab+\tilde \thetab} }{\partial{\phi_k}} \Big]\\
% &=\mathbb{E}_{\tilde{\theta}}\Big[\frac{\expval{O_a}_{o, \phis + c_k\mathbf{e}_k, \thetab} + \expval{O_a}_{o, \phis - c_k\mathbf{e}_k, \thetab}}{2}\cdot \frac{\expval{O_a}_{o,\phis + c_k\mathbf{e}_k ,\thetab+\tilde \thetab} -\expval{O_a}_{o,\phis - c_k\mathbf{e}_k,\thetab}}{2|c_k|}, \Big]
% \end{align}
% \begin{multline}
% \mathbb{E}_{\tilde{\theta}}\left[\frac{\partial \mathcal{L}(\phis;\thetab+\tilde{\thetab})}{\partial{\phi_k}} \right]= \frac{\sin\alpha}{\alpha}\cdot\frac{\partial \expval{O_a}_{o,\phis,\thetab} }{\partial{\phi_k}}\cdot \\ 
% \frac{\beta^2}{|c_k| n(\mathcal{E})} \sum_{\expval{o,a,r,o'} \in \mathcal{E}}  \left(\frac{r}{\beta}+\max_{a'}\expval{O_{a'}}_{o', \phis', \thetab} - \frac{\expval{O_a}_{o, \phis + c_k\mathbf{e}_k, \thetab} + \expval{O_a}_{o, \phis - c_k\mathbf{e}_k, \thetab}}{2}\cdot  \frac{\sin\alpha}{\alpha}\right).
% \end{multline}
% Summarizing with $\frac{\partial \mathcal{L}(\phis;\thetab)}{\partial{\phi_k}}$, $A_1$, the formula is as written as \eqref{eq:lem2}.
\end{proof}

\subsection{Lemma 3: \textit{Variance of meta Q-network's gradient with angle noise regularizer}}\label{app:proof_lem_3}
\begin{lemma}\label{lem:lemma3}
The variance of meta \textit{Q}-network gradient with angle regularizer is bounded to a constant written as follows: \begin{equation}\label{eq:lem3-3}
         \mathrm{Var}\left[\frac{\partial \mathcal{L}(\phis;\thetab+\tilde{\thetab}) }{\partial{\phi_k}}\right] \leq  \frac{4\beta^4}{|\mathcal{E}|^2}\sum_{\tau\in\mathcal{E}}\Big(\frac{\sin2\alpha}{2\alpha}A_3(\tau)\cdot\mathrm{Tr}(A_4^2M^2_{\theta_m})  - \frac{\sin^2\alpha}{\alpha^2}\mathrm{Tr}(A_4M_{\theta_m})^2\Big),
     \end{equation}
     where $A_3(\tau)={\Big(\frac{r}{\beta}+\max_{a'}\expval{O_{a'}}_{o', \phis', \thetab} - 1 \Big)^2}$, $A_4 = \ket{\ppsi}\bra{\ppsi}-\ket{\mpsi}\bra{\mpsi}$.
\end{lemma}
% \subsection{Proof of Corollary 1}\label{app:proof_col_1}
% \begin{proof}
%  Substituting $\frac{\partial \expval{O_a}_{o,\phis,\thetab} }{\partial{\phi_k}}$ into $\frac{\partial \mathcal{L}(\phis;\thetab) }{\partial{\phi_k}} $, we derive the equation above.
% \end{proof}
\begin{proof}
The variance of meta \textit{Q}-network is written as follows:
\begin{equation}
\mathrm{Var}_{\tilde{\thetab}}\left[\frac{\partial \mathcal{L}(\phis;\thetab+\tilde{\thetab}, \mathcal{E}) }{\partial{\phi_k}}\right] =\mathbb{E}_{\tilde \thetab}\Bigg[\underbrace{\frac{\partial \mathcal{L}(\phis;\thetab+\tilde{\thetab}, \mathcal{E}) }{\partial{\phi_k}}^2}_{\circled{3}}\Bigg]- \mathbb{E}_{\tilde{\thetab}}\Bigg[\underbrace{\Big(\frac{\partial \mathcal{L}(\phis;\thetab+\tilde{\thetab}, \mathcal{E}) }{\partial{\phi_k}}\Big)^2}_{\circled{4}}\Bigg].
\end{equation}
$\circled{3}$ can be written as follows:
\begin{equation}
    \circled{3} = \frac{4\beta^4}{|\mathcal{E}|^2}\Bigg(\sum_{\expval{o,a,r,o'} \in \mathcal{E}} \Big(\frac{r}{\beta}+\max_{a'}\expval{O_{a'}}_{o', \phis', \thetab} - \expval{O_a}_{o,\phis,\thetab+\tilde \thetab} \Big) \cdot \frac{\partial \expval{O_a}_{o,\phis,\thetab+\tilde \thetab} }{\partial{\phi_k}}\Bigg)^2.
\end{equation}
By the triangle inequality, $\circled{3}$ is bounded to the summation of the square as follows: 
\begin{equation}
     \circled{3} \leq \frac{4\beta^4}{|\mathcal{E}|^2}\sum_{\expval{o,a,r,o'} \in \mathcal{E}} \Big(\frac{r}{\beta}+\max_{a'}\expval{O_{a'}}_{o', \phis', \thetab} - \expval{O_a}_{o,\phis,\thetab+\tilde \thetab} \Big)^2 \cdot \Big(\frac{\partial \expval{O_a}_{o,\phis,\thetab+\tilde \thetab} }{\partial{\phi_k}}\Big)^2.
\end{equation}Since, $\expval{O_a}_{o,\phis,\thetab+\tilde \thetab} \leq 1$, $\mathbb{E}_{\tilde{\theta}}[\circled{3}]$ is written as follows:
\begin{equation}\label{eq:lem3-0-3}
    \mathbb{E}_{\tilde{\theta}_i}\Big[\circled{3}\Big] \leq \frac{4\beta^4}{|\mathcal{E}|^2}\sum_{\expval{o,a,r,o'} \in \mathcal{E}} \underbrace{\Big(\frac{r}{\beta}+\max_{a'}\expval{O_{a'}}_{o', \phis', \thetab} - 1 \Big)^2}_{A_3(\tau)} \cdot\underbrace{ \mathbb{E}_{\tilde{\theta}}\Big[\Big(\frac{\partial \expval{O_a}_{o,\phis,\thetab+\tilde \thetab} }{\partial{\phi_k}}\Big)^2\Big]}_{\circled{3a}}.
\end{equation}

Now, we focus on $\circled{3a}$. To handle $\circled{3a}$, we use trace operation denoted as $\mathrm{Tr}(\cdot)$. For simplicity, we denote the quantum state of $\expval{O_a}_{o,\phis + c_k\mathbf{e}_k ,\thetab}$ and  $\expval{O_a}_{o,\phis - c_k\mathbf{e}_k ,\thetab}$  as $\psi^+$,  $\psi^-$, respectively. Then, the partial derivative is written as follows:
     \begin{equation}
     \frac{\partial \expval{O}_{o,\phis;\thetab+\tilde{\thetab}}}{\partial{\phi_k}}=\mathrm{Tr}((\ket{\ppsi}\bra{\ppsi}-\ket{\mpsi}\bra{\mpsi})M_{\theta_m}M_{\theta_m + \tilde{\theta}_m}Z).
     \end{equation}
     Let's denote $A_4 = \ket{\ppsi}\bra{\ppsi}-\ket{\mpsi}\bra{\mpsi}$. Then, the expectation of squared partial derivative is written as follows: 
     \begin{equation}\label{eq:lem3-222}
         \mathbb{E}_{\tilde{\thetab}}\Big[\Big(\frac{\partial \expval{O}_{\phis;\thetab+\tilde{\thetab}}}{\partial{\phi_k}}\Big)^2\Big] = \mathrm{Tr}(A_4^2M_{\theta_m}^2\mathbb{E}_{\tilde{\theta}_i}[M_{\theta_m + \tilde{\theta}_m}^2]Z^2)=\frac{\sin2\alpha}{2\alpha}\mathrm{Tr}(A_4^2M_{\theta_m}^2),
     \end{equation}
     Substituting the RHS of \eqref{eq:lem3-222} into \eqref{eq:lem3-0-3}, we derive the bound of $\circled{3}$ as follows:
     \begin{equation}
         \circled{3} \leq \frac{4\beta^4}{|\mathcal{E}|^2}\frac{\sin2\alpha}{2\alpha}\sum_{\tau\in\mathcal{E}}A_3(\tau)\cdot\text{Tr}(A_4^2M^2_{\theta_m})
     \end{equation}
     Now, we focus on figuring out $\circled{4}$. The expectation of partial derivative is rewritten as,
     \begin{equation}\label{eq:lem3-2}
         \mathbb{E}_{\tilde{\theta}_i}\Big[\frac{\partial \expval{O}_{\phis;o, \thetab+\tilde{\thetab}}}{\partial{\phi_k}}\Big] = \mathrm{Tr}(A_4M_{\theta_m}\mathbb{E}_{\tilde{\theta}_i}[M_{\theta_m + \tilde{\theta}_m}]Z)=\frac{\sin\alpha}{\alpha}\mathrm{Tr}(A_4M_{\theta_m}).
     \end{equation}
Then the squared of expectation can be written as follows: 
\begin{equation}
    \mathbb{E}_{\tilde \thetab}[\circled{4}] =   \frac{4\beta^4}{|\mathcal{E}|^2}\frac{\sin^2\alpha}{\alpha^2}\Big({\sum_{\tau \in \mathcal{E}} A_2(\tau) \frac{\partial \expval{O_a}_{o,\phis,\thetab} }{\partial{\phi_k}}}\Big)^2.
\end{equation}
Finally, the gradient variance is summarized as follows:
     \begin{equation}
         \mathrm{Var}\left[\frac{\partial \mathcal{L}(\phis;\thetab+\tilde{\thetab}) }{\partial{\phi_k}}\right] \leq  \frac{4\beta^4}{|\mathcal{E}|^2}\sum_{\tau\in\mathcal{E}}\Big(\frac{\sin2\alpha}{2\alpha}A_3(\tau)\cdot\text{Tr}(A_4^2M^2_{\theta_m})  - \frac{\sin^2\alpha}{\alpha^2}\mathrm{Tr}(A_4M_{\theta_m})^2\Big),
     \end{equation}
and this result finalizes the proof. \end{proof}
\subsection{Proof of Theorem 1}\label{app:proof_thm_1}
\begin{proof}
Denote $\tilde{\phis}_i$ as a meta \textit{Q}-network parameter at the epoch $i$ with angle noise. For the simplicity, we denote $\tau = \langle o, a, r, o'\rangle$. From Lemma~\ref{lem:lemma3}, we have gradient variance has a constant bound, then, we have following derived from Lemma ~\ref{lem:lemma2}:

\begin{align}
         &\mathbb{E}_{\tilde{\thetab}}\norm{\tilde\phis_i-\phis^*} \nonumber \\
         &= \mathbb{E}_{\tilde{\thetab}}\norm{\sum_{j=i}^{\infty}\eta_i\nabla_{\phis}\mathcal{L}(\phis,\thetab+\tilde\thetab;\mathcal{E}_j)}
         = \norm{\sum_{j=i}^{\infty}\eta_j\mathbb{E}_{\tilde{\thetab}}[\nabla_{\phis}\mathcal{L}(\phis,\thetab+\tilde\thetab;\mathcal{E}_j)]} \nonumber \\
         &= 
        \norm{\sum_{j=i}^{\infty}\frac{2\beta^2\eta_j}{|\mathcal{E}_j|}\frac{\sin\alpha}{\alpha}\sum_{\tau \in \mathcal{E}_j}A_2(\tau)\cdot \nabla_{\phis}\expval{O_a}_{o,\phis,\thetab}} \nonumber \\
        &\leq \norm{\sum_{j=i}^{\infty}\frac{2\beta^2\eta_j}{|\mathcal{E}_j|}\frac{\sin\alpha}{\alpha}\sum_{\tau \in \mathcal{E}_j}
        A_1\cdot \nabla_{\phis}\expval{O_a}_{o,\phis,\thetab}}+ \norm{\sum_{j=i}^{\infty}\frac{2\beta^2\eta_j}{|\mathcal{E}_j|}\frac{\sin\alpha}{\alpha}\sum_{\tau \in \mathcal{E}_j} \nabla_{\phis}\expval{O_a}_{o,\phis,\thetab}} \nonumber  \\
        &\leq
        \frac{\sin\alpha}{\alpha}\underbrace{\norm{\sum_{j=i}^{\infty}\eta_j\nabla_{\phis}\mathcal{L}(\phis,\thetab;\tau)}}_ {\circled{5}} + \frac{\sin\alpha}{\alpha}\underbrace{\norm{\sum_{j=i}^{\infty}\frac{2\beta^2\eta_j}{|\mathcal{E}_j|}\sum_{\tau\in\mathcal{E}_j}\nabla_{\phis} \expval{O}_{o, \phis, \thetab}}}_{\circled{6}}
\end{align}
Since meta \textit{Q}-network follows bi-Lipschitz continuous, $\circled{5}$ is bounded to $\epsilon_{i}$. 
According to the result of Lemma 3, the variance of meta \textit{Q}-network's gradient is bounded to constant. Then $\circled{6}$ is bound to constant, we prove that \textit{meta \textit{Q}-network with noise regularizer has a convergence bound}.
%Note that $D_1$ is bounded to $\epsilon_i$. If meta \textit{Q}-network converges, \textit{i.e.,} $D_2 \leq \epsilon_i'$, meta \textit{Q}-network with quantum angle noise also converges. 
\end{proof}
\section{Simulation Setting} \label{app:simul}
\begin{table}[h!]
\centering
\caption{Common experimental parameters}
\begin{tabular}{lc}
\toprule[1pt]
    \multicolumn{1}{c}{\textbf{Paramters}} & \textbf{Values} \\
    \midrule
    Learning rate ($\eta_0)$ & $10^{-4}$\\
    Weight decay & $10^{-5}$\\
    Number of angle parameters $(|\phis|)$ & 45\\
    Number of measurement parameters$(|\thetab|)$ & 6\\
    Hyperparameter of observable $(\beta)$ & 8\\
    Optimizer  & Adam Optimizer \\
    Angle noise bound $(\alpha)$ & $\{0, \pi/6, \pi/4, \pi/3, \pi/2\}$\\
     \bottomrule[1pt]
\end{tabular}
\label{tab:setting}  
\end{table}
The simulation setting is presented in Table \ref{tab:setting}. We train meta-\textit{Q} network with $2,000$ training epochs. Then, we conduct few-shot angle training with $20,000$ training epochs. We pre-train the policy with the angle noise $\alpha = \{0, 30^\circ, 45^\circ, 60^\circ,90^\circ\}$. To enhance the performance, we design the measurement composing two rotation gates for each qubit to leverage the full potential of Bloch sphere expressibility. Then, agents' measurement parameters $\Theta =\{\thetab^j\}^N_{j=1}$ are trained with CTDE method. We train in a two-step game and a single-hop offloading environment. For the simulation, we use \texttt{Pytorch v1.8.2 LTS}, \texttt{Torchquantum} as software libraries. The experiments are conducted on \texttt{NVIDIA DGX Station V100 (4EA)}. For reproductivity, we provide experiment codes in the supplemental material and all source codes follow the \texttt{MIT License}.

\section{Two-Step Game} \label{app:twostep}
\begin{figure}[ht!]
\label{fig:twostep}
    \centering
        \includegraphics[width=0.6\textwidth]{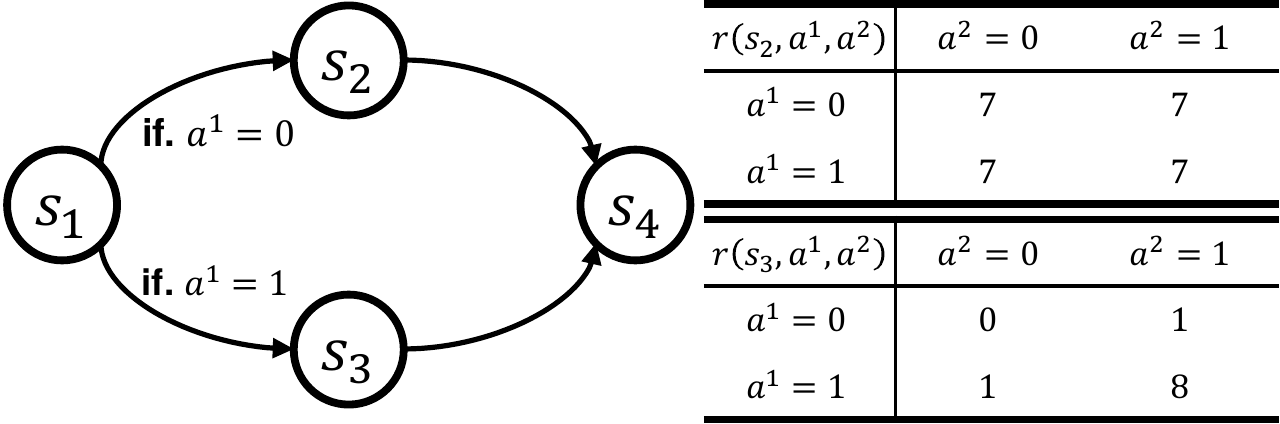}
    \caption{The state diagram of prisoner's dilemma and corresponding reward table.}
    \label{fig:prisoner}
\end{figure}
Fig.~\ref{fig:prisoner} gives the simple case of a two-agent matrix game. It starts at $s_1$. The state transition from $s_1$ is determined by an action of the first agent, regardless of the action of the second agent. In state $s_2$, all agents receive the score $7$. In state $s_3$, an agent receives a reward that depends on the other agent's action. 

If agent does not know the other agent's strategy, the optimal action value function is $\mathbb{E}[\max_aQ^*(s_1, a)] = 7$, $\mathbb{E}[\max_aQ^*(s_2, a)] = 7$, and $\mathbb{E}[\max_aQ^*(s_3, a)] = 4.5$, respectively. If agent knows the the other agent's strategy, the optimal action value function $\mathbb{E}[\max_aQ^*(s_1, a)] = 8$, $\mathbb{E}[\max_aQ^*(s_2, a)] = 7$, and $\mathbb{E}[\max_aQ^*(s_3, a)] = 8$, respectively.
\section{Continual Learning}\label{app:continual}
\begin{table}[ht]\resizebox{1\linewidth}{!}{
\begin{tabular}{ccc}
\begin{tabular}{c|cc}\toprule[2pt]
    $r(s_1,a^1,a^2)$ &  $a^2=0$ & $a^2=1$ \\\midrule
    $a^1=0$          &  0       & 0       \\
    $a^1=1$          &  0       & 0       \\\bottomrule[2pt]
\end{tabular} &
\begin{tabular}{c|cc}\toprule[2pt]
    $r(s_2,a^1,a^2)$ &  $a^2=0$ & $a^2=1$ \\\midrule
    $a^1=0$          &  4       & 4       \\
    $a^1=1$          &  4       & 4       \\\bottomrule[2pt]
\end{tabular} &
\begin{tabular}{c|cc}\toprule[2pt]
    $r(s_3,a^1,a^2)$ &  $a^2=0$ & $a^2=1$ \\\midrule
    $a^1=0$          &  0       & 1       \\
    $a^1=1$          &  1       & 8       \\\bottomrule[2pt]
\end{tabular}\\[5pt]
\end{tabular}}\\[5pt]
\caption{The reward function of environment A.}
\resizebox{1\linewidth}{!}{
\begin{tabular}{ccc}
\begin{tabular}{c|cc}\toprule[2pt]
    $r(s_1,a^1,a^2)$ &  $a^2=0$ & $a^2=1$ \\\midrule
    $a^1=0$          &  0       & 0       \\
    $a^1=1$          &  0       & 0       \\\bottomrule[2pt]
\end{tabular} &
\begin{tabular}{c|cc}\toprule[2pt]
    $r(s_2,a^1,a^2)$ &  $a^2=0$ & $a^2=1$ \\\midrule
    $a^1=0$          &  4       & 4       \\
    $a^1=1$          &  4       & 4       \\\bottomrule[2pt]
\end{tabular} &
\begin{tabular}{c|cc}\toprule[2pt]
    $r(s_3,a^1,a^2)$ &  $a^2=0$ & $a^2=1$ \\\midrule
    $a^1=0$          &  8       & 1       \\
    $a^1=1$          &  1       & 1       \\\bottomrule[2pt]
\end{tabular}\\[5pt]
\end{tabular}}\\[5pt]
\caption{The reward function of environment B.}
\end{table}
\BfPara{Environment design for continual learning}
We additionally design a couple of environments to conduct experiments for continual learning. 
In continual learning, the environment is changed (A$\rightarrow$B$\rightarrow$A). In environment A, when multiple agents are in $s_3$, the optimal action is $\mathbf{a}=1$. In environment B, when multiple agents are in $s_3$, the optimal action is $\mathbf{a}=0$.
\section{Single-hop Offloading Environment}\label{app:singlehop}
\begin{table}[ht!]
\centering
\caption{The MDP of a single-hop offloading environment.}
\begin{tabular}{p{0.2\textwidth}|p{0.6\textwidth}}
    \toprule[1pt]
    \textbf{Component} & \textbf{Description}\\\midrule\midrule
    \textbf{Number of agents}& $4$.\\\midrule
    \textbf{Observation}& Current and previous queue state of agent and current queue state of clouds, $\dim(o) = 4$.\\\midrule
    \textbf{State} & $s_t \triangleq \cup^N_{n=1}\{o^n_t\}$, $\dim(s) = 16$.\\\midrule
    \textbf{Action}& Transmit small/large amount of chunk to clouds, $\dim(a) = $Discrete($4$).\\\midrule
    \textbf{Reward} & Modeled with queue stability. \\
    \midrule\textbf{Expected returns} & $-33.2$ with random policy.\\\bottomrule[1pt]
\end{tabular}
\label{tab:mdp}
\end{table}
\citet{yun2022QMARL} have proposed the single-hop environments for quantum multi-agent reinforcement learning environment. The single-hop environments consist of 4 edge agents and 2 cloud agents. A queue is given for every edge agent and every cloud, where the queue can temporally store some tasks (\textit{e.g.,} cache, or chunk). The objective of this environment is to maintain queue stability for all agents and clouds. The edge agents make action decisions of which the action space consists of the amount of transmitting chunk and the destination cloud. The MDP formulation is presented in Table~\ref{tab:mdp}.

\BfPara{Simulation Settings} Experiments are conducted with four different schemes; 1) a technique that removes the meta \textit{Q}-network training, 2) QM2ARL but removes the angle noise (\textit{i.e.,} $\alpha= 0^\circ$), 3) a state-of-the-art CTDE QMARL method~\cite{yun2022QMARL}, 4) Learning about total reward in the QM2ARL and angle noise method proposed in this paper (\textit{i.e.,} $\alpha= 30^\circ$). 

\section{Additional Results}\label{app:additional_result}

\subsection{Results of Local-QNN Pole Training}\label{app:additional_result1}
\begin{figure*}[ht!]
    \centering
    \includegraphics[width=.95\linewidth]{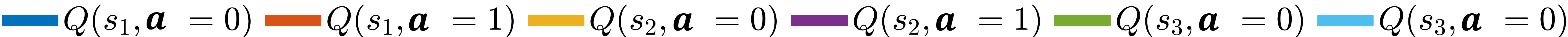}
    \begin{tabular}{p{0.17\linewidth}p{0.17\linewidth}p{0.17\linewidth}p{0.17\linewidth}p{0.17\linewidth}}
        \includegraphics[width=\linewidth]{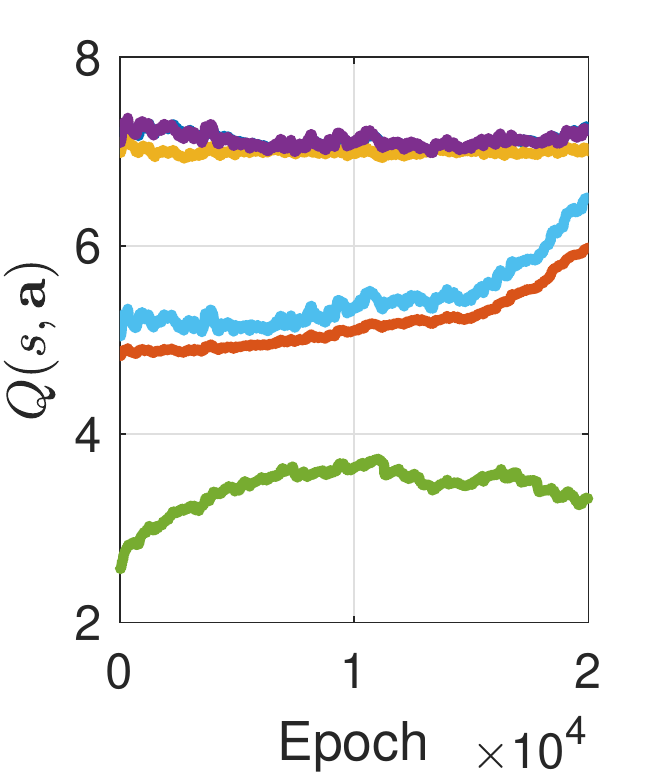}
        & \includegraphics[width=\linewidth]{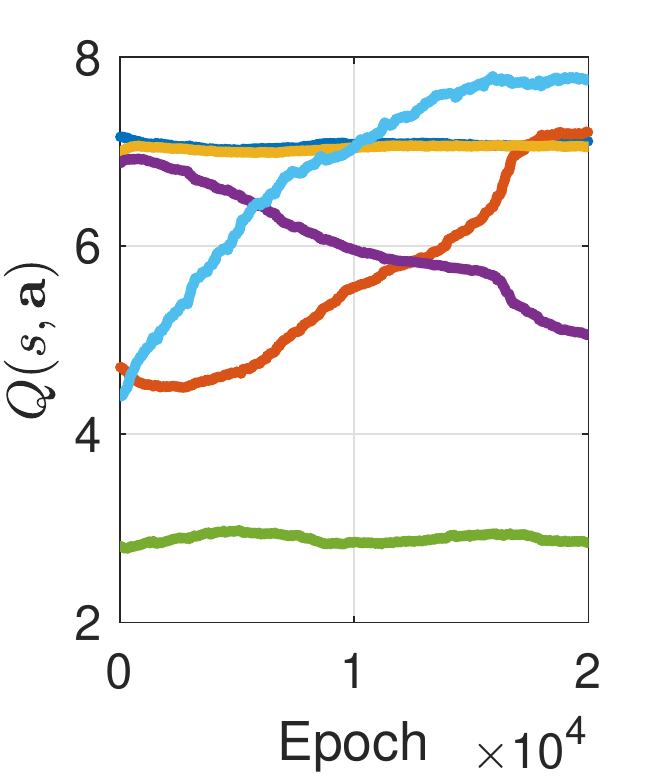}
        & \includegraphics[width=\linewidth]{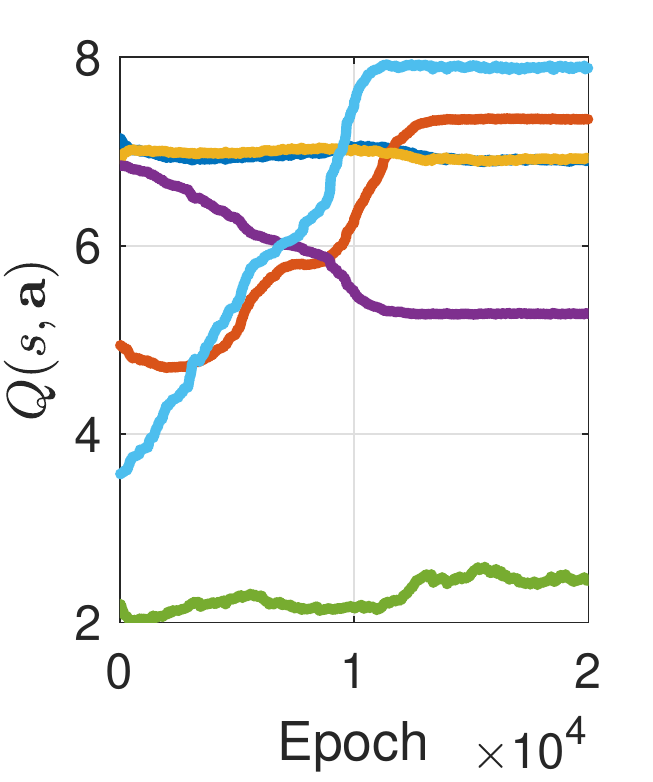}
        & \includegraphics[width=\linewidth]{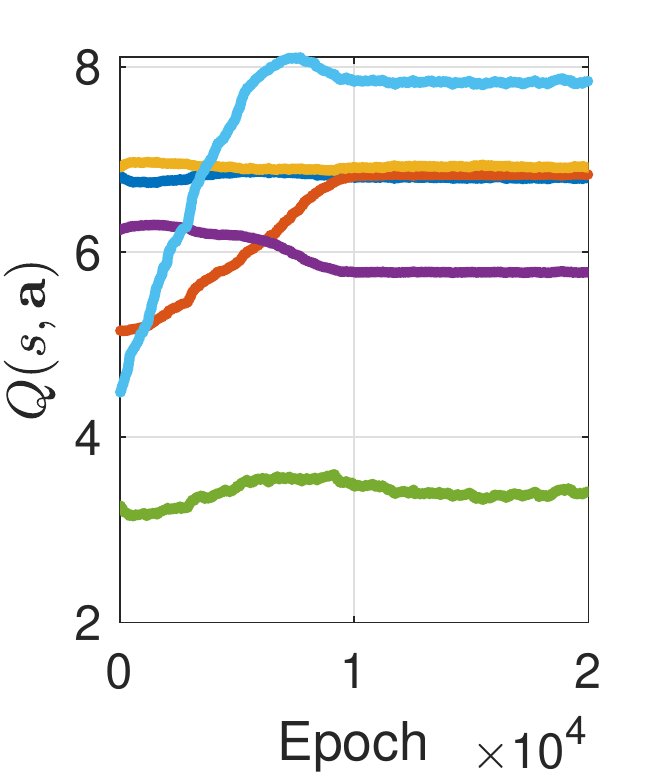}
        & \includegraphics[width=\linewidth]{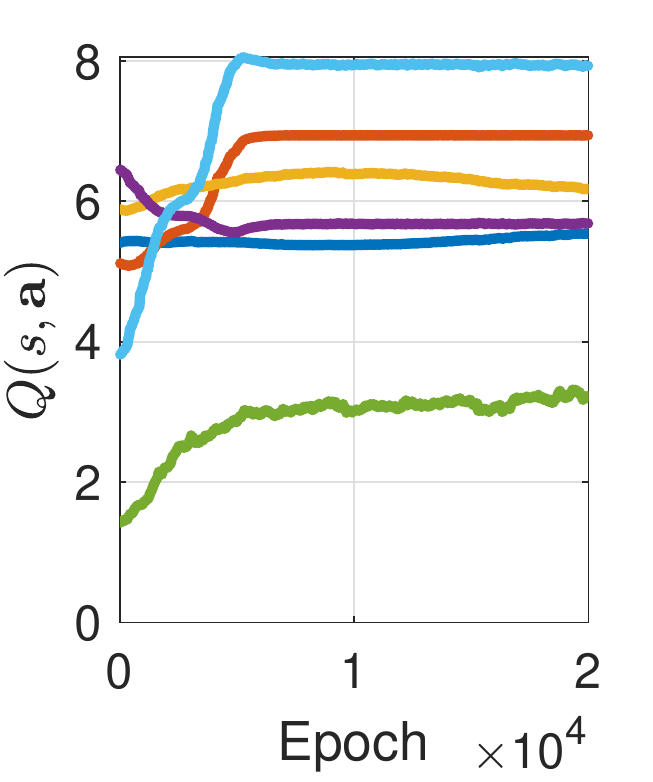}\\
        \multicolumn{1}{c}{  \small (a) $\alpha=0^\circ$} 
        & \multicolumn{1}{c}{\small (b) $\alpha=30^\circ$} 
        & \multicolumn{1}{c}{\small (c) $\alpha=45^\circ$} 
        & \multicolumn{1}{c}{\small (d) $\alpha=60^\circ$} 
        & \multicolumn{1}{c}{\small (e) $\alpha=90^\circ$}\\
    \end{tabular}
    \caption{The tendency of the action value function in local-QNN pole training; (a--e) shows action value difference corresponding to the different angle noise regularizers.}
    \label{fig:mt}
\end{figure*}

\subsection{Results of Meta \textit{Q}-Network Training}\label{app:additional_result2}

\begin{figure*}[t!]
    \centering
\begin{tabular}{cp{0.158\linewidth}p{0.158\linewidth}p{0.158\linewidth}p{0.158\linewidth}p{0.158\linewidth}}
& \multicolumn{1}{c}{\small $\alpha=0^\circ$} 
& \multicolumn{1}{c}{\small $\alpha=30^\circ$} 
& \multicolumn{1}{c}{\small $\alpha=45^\circ$} 
& \multicolumn{1}{c}{\small $\alpha=60^\circ$} 
& \multicolumn{1}{c}{\small $\alpha=90^\circ$}\\
\rotatebox{90}{\small $D_{\text{norm}}(s_1,\phis,\thetab)$} 
& \includegraphics[width=\linewidth]{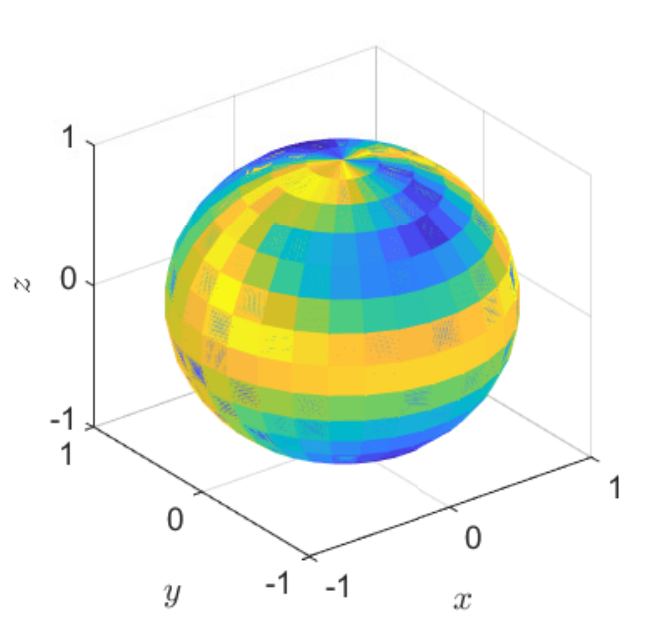}
& \includegraphics[width=\linewidth]{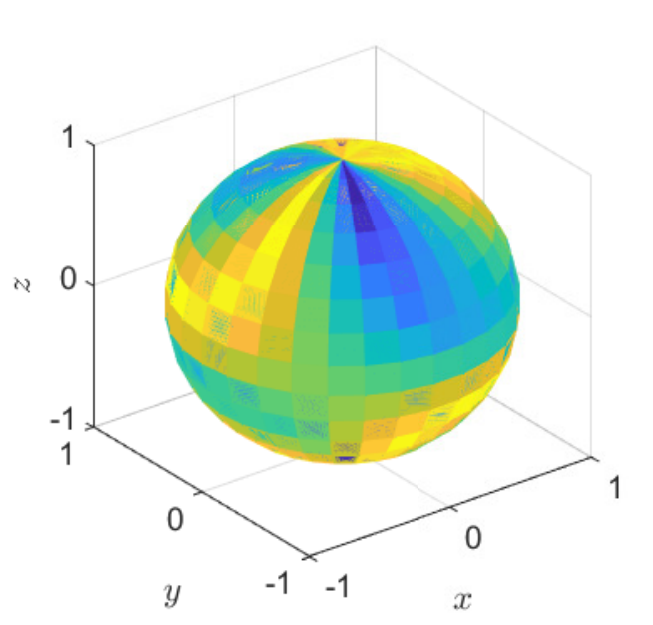}
& \includegraphics[width=\linewidth]{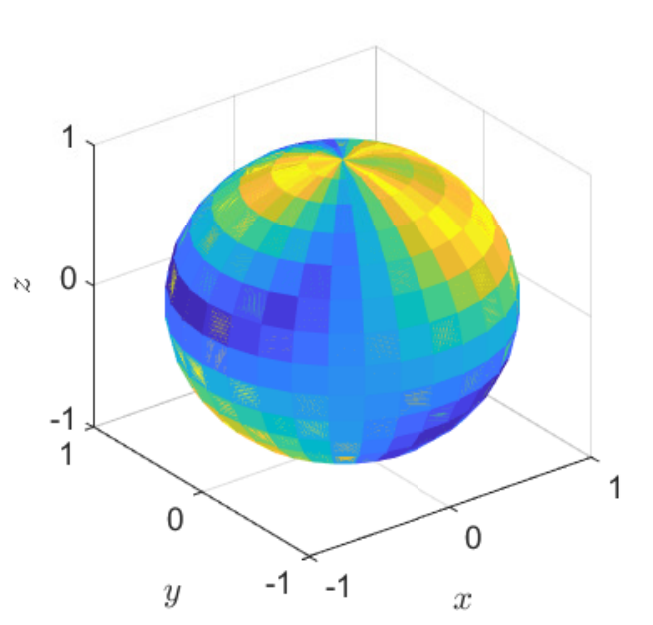}
& \includegraphics[width=\linewidth]{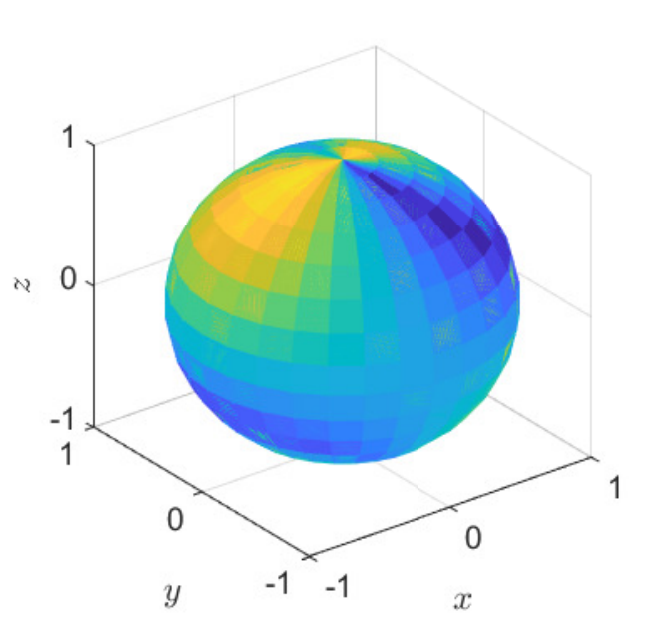}
& \includegraphics[width=\linewidth]{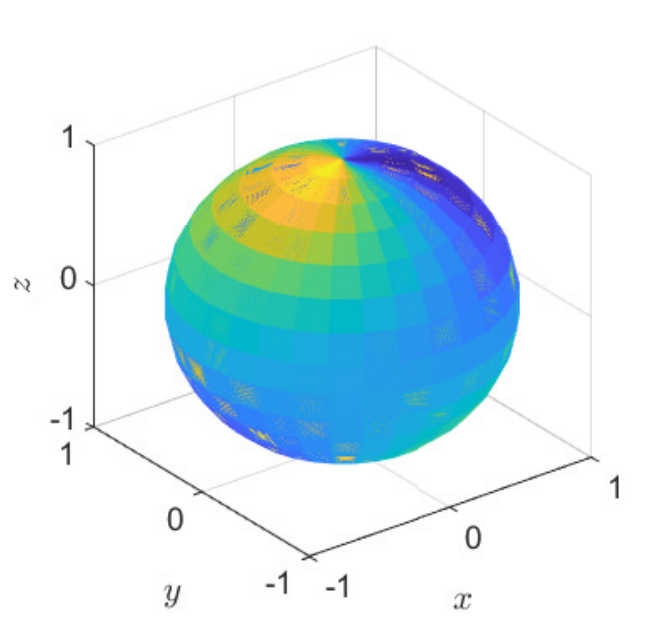}\\
\rotatebox{90}{\small $D_{\text{norm}}(s_2,\phis,\thetab)$} 
& \includegraphics[width=\linewidth]{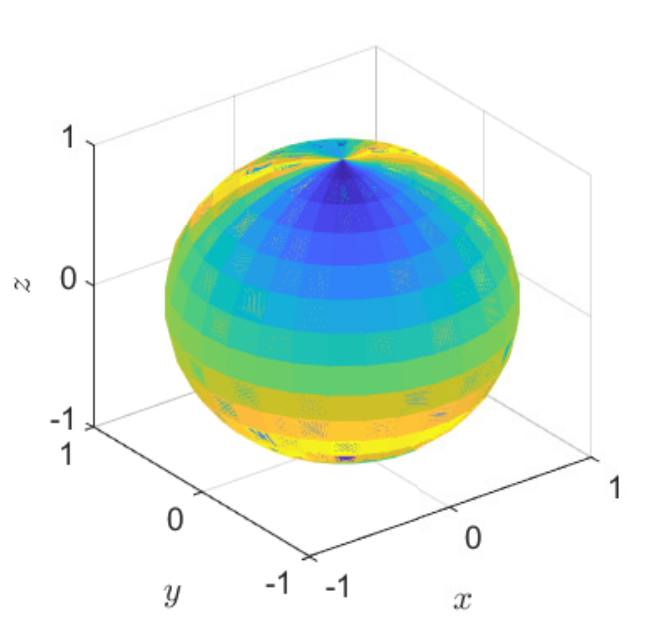}
& \includegraphics[width=\linewidth]{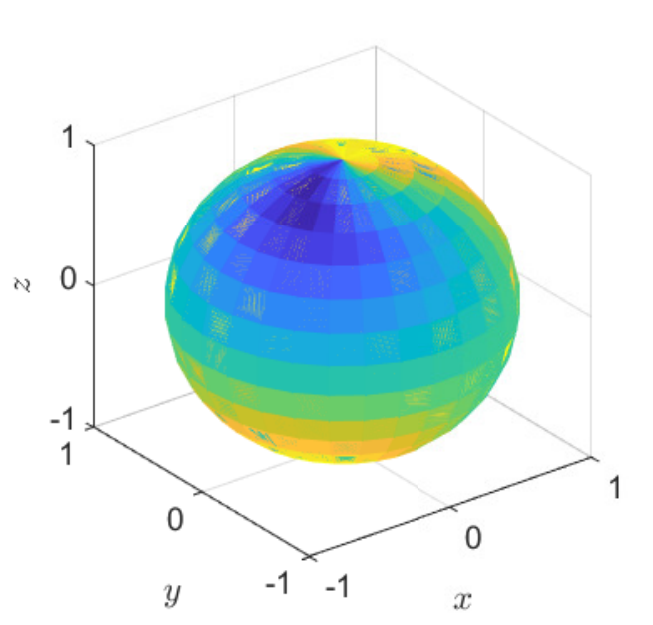}
& \includegraphics[width=\linewidth]{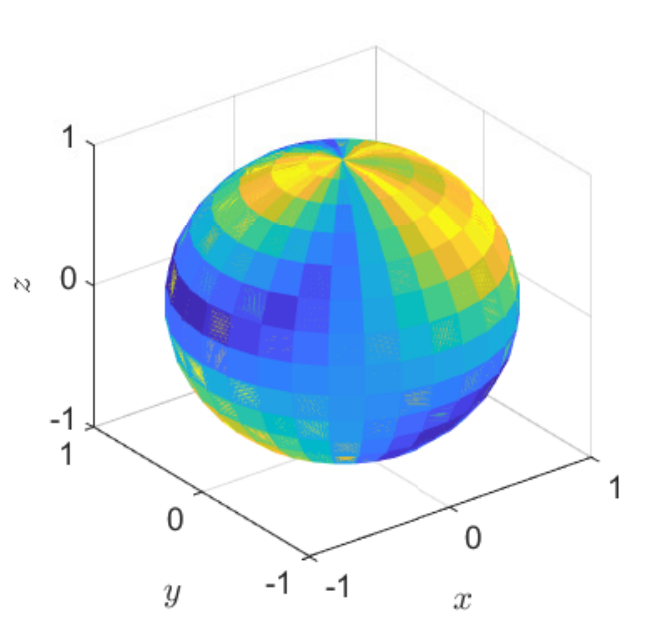}
& \includegraphics[width=\linewidth]{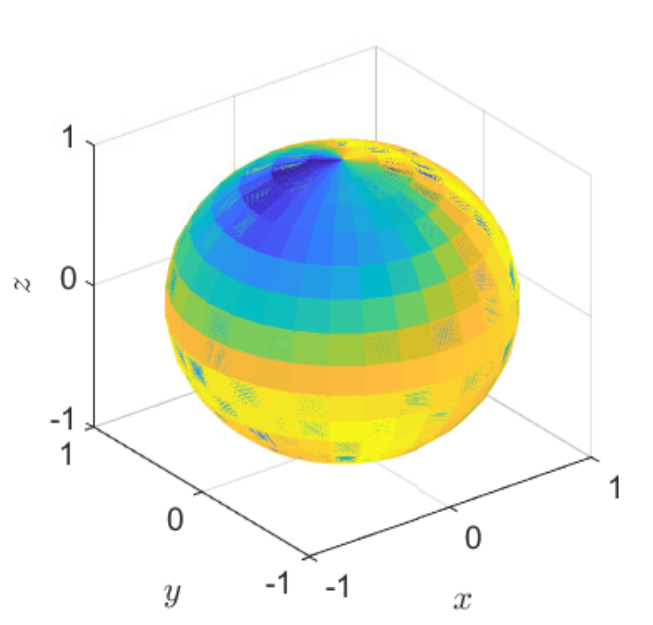}
& \includegraphics[width=\linewidth]{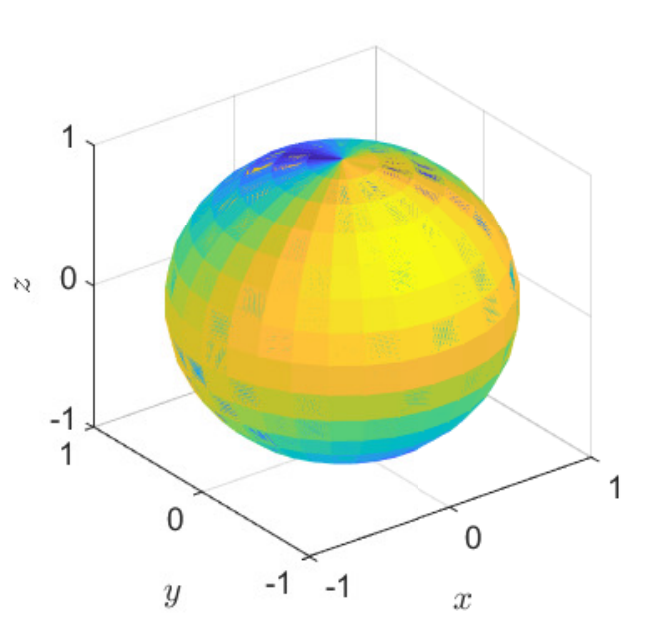}\\
\rotatebox{90}{\small $D_{\text{norm}}(s_3,\phis,\thetab)$} 
& \includegraphics[width=\linewidth]{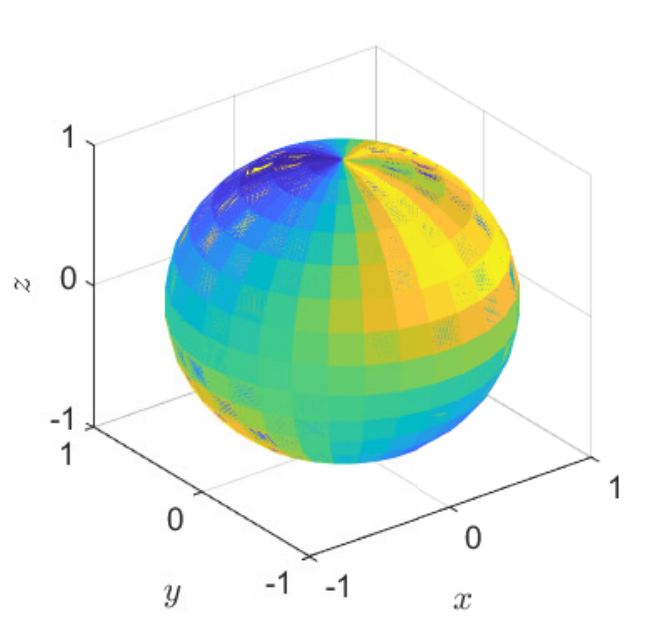}
& \includegraphics[width=\linewidth]{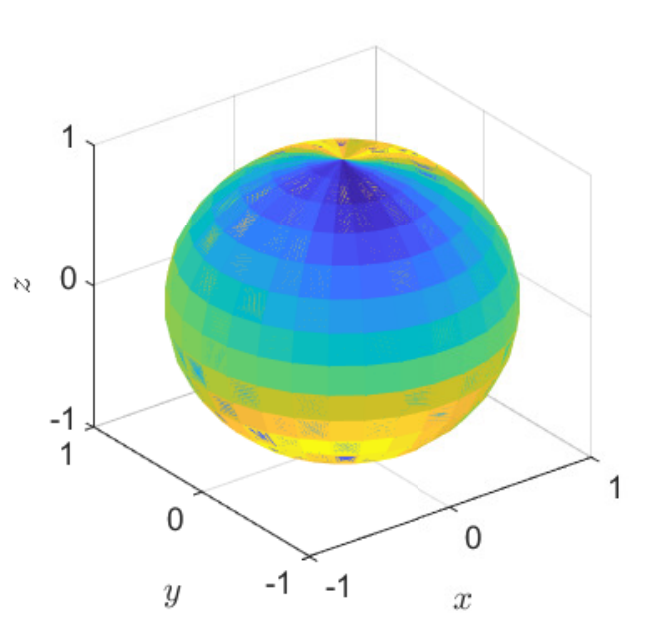}
& \includegraphics[width=\linewidth]{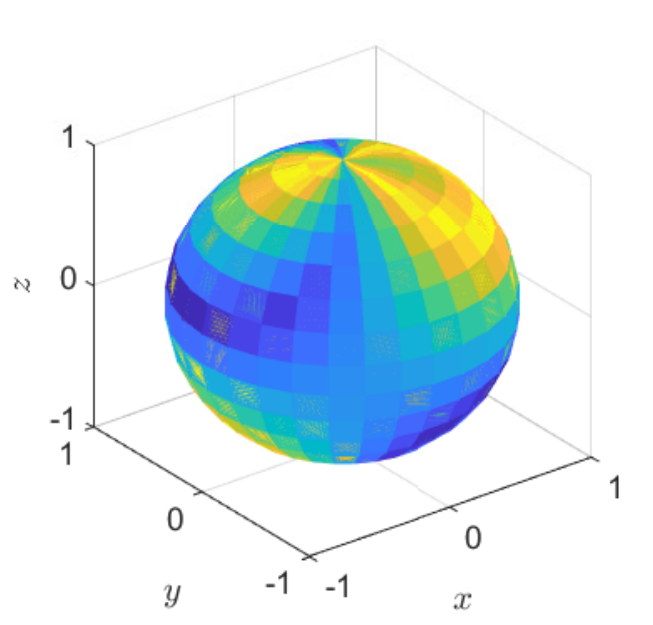}
& \includegraphics[width=\linewidth]{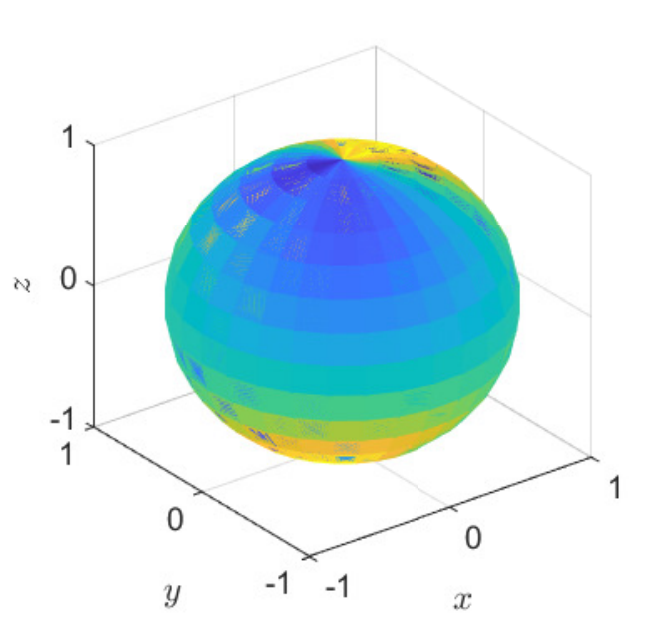}
& \includegraphics[width=\linewidth]{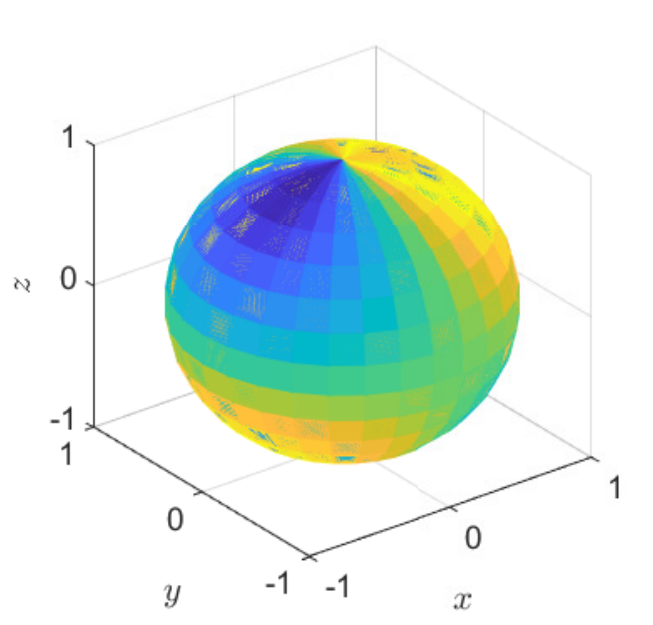}\\
\end{tabular}
    \caption{Comparison of distance between meta-agent's \textit{Q}-value and optimal action value corresponding to the poles.}
    \label{fig:bloch_pretrain}
\end{figure*}

\begin{table}[ht]
\centering
\small
\begin{tabular}{ccc}
\begin{tabular}{c|cc}\toprule[2pt]
    $Q(s,a)$ &  $a=0$ & $a=1$ \\\midrule
    $s_1$          &  6.916    &4.399    \\
    $s_2$          &  6.965   & 6.932   \\
    $s_3$          &  2.893   & 4.163   \\\bottomrule[2pt]
\end{tabular} &
\begin{tabular}{c|cc}\toprule[2pt]
    $Q(s,a)$ &  $a=0$ & $a=1$ \\\midrule
    $s_1$          &  6.955   & 4.416   \\
    $s_2$          &  7.109   & 6.458   \\
    $s_3$          &  2.908   & 3.168   \\\bottomrule[2pt]
\end{tabular}
 &
\begin{tabular}{c|cc}\toprule[2pt]
    $Q(s,a)$ &  $a=0$ & $a=1$ \\\midrule
    $s_1$          &  7.352   & 4.692   \\
    $s_2$          &  5.915   & 6.611   \\
    $s_3$          &  4.262   & 3.810   \\\bottomrule[2pt]
\end{tabular} 
\\
\\
(a) $\alpha = 0^\circ$. &
(b) $\alpha = 30^\circ$. &
(c) $\alpha = 45^\circ$. \\
\\
\end{tabular}\\
\begin{tabular}{ccc}
\begin{tabular}{c|cc}\toprule[2pt]
    $Q(s,a)$ &  $a=0$ & $a=1$ \\\midrule
    $s_1$          &  6.972   & 5.785   \\
    $s_2$          &  6.015   & 6.087   \\
    $s_3$          &  4.585   & 6.814   \\\bottomrule[2pt]
\end{tabular} &
\begin{tabular}{c|cc}\toprule[2pt]
    $Q(s,a)$ &  $a=0$ & $a=1$ \\\midrule
    $s_1$          &  5.634   & 6.267  \\
    $s_2$          &  6.651   & 6.498   \\
    $s_3$          &  4.128   & 4.223   \\\bottomrule[2pt]
\end{tabular} &
\begin{tabular}{c|cc}\toprule[2pt]
    $Q(s,a)$ &  $a=0$ & $a=1$ \\\midrule
    $s_1$          &  7       & 4.5     \\
    $s_2$          &  7       & 7       \\
    $s_3$          &  0.5     & 4.5     \\\bottomrule[2pt]
\end{tabular}
\\
\\
(d) $\alpha = 60^\circ$. &
(e) $\alpha = 90^\circ$. &
(f) Ground Truth. \\
\end{tabular}
\vspace{10pt}
    \caption{The results of meta \textit{Q}-network training process (test \textit{Q}-table).}
    \label{fig:test_qtable}
\end{table}
Table~\ref{fig:test_qtable} indicates the test \textit{Q}-value. %However, it is not intuitive to interpret the result.
To verify the impact of angle regularizer on the expressibility of meta-\textit{Q} network, we design the benchmark scheme by checking how much area on the Bloch sphere indicating optimal values occupies. We consider the comparisons as meta \textit{Q}-networks with various angle noise bounds, \textit{i.e.,} $\alpha = \{0^\circ, 30^\circ, 45^\circ, 60^\circ, 90^\circ\}$. 
For this, we calculate the distance between an optimal action-value and the action-value of meta-\textit{Q} network, where the same state is given. Then we min-max normalize the distance. The process is written as follows:
\begin{eqnarray}\label{eq:distance}
D(s, \phis, \thetab) &=& \norm{\max_a Q^*(s,a) -  \max_a Q(s,a;\phis,\thetab)},\\
D_{\text{norm}}(s, \phis, \thetab) &=& \frac{D(s, \phis, \thetab)}{\max_{\thetab}(D(s, \phis, \thetab) )- \min_{\thetab}(D(s, \phis, \thetab) )}.
\end{eqnarray}
We probe meta Q-values $\max_a Q(s,a;\phis,\thetab)$ for all the measurement parameters that are quantized to $\thetab \in \{ -\frac{16 \pi}{16},-\frac{15 \pi}{16}, \cdots, 0, \cdots, \frac{15 \pi}{16}, \frac{16 \pi}{16},\}^{\abs{\thetab}}$. 
Fig.~\ref{fig:bloch_pretrain} presents the $D_{\text{norm}}$ on the Bloch sphere for every qubit. As the surface color is more \textit{blue}, the action value is far from the optimal action value function. In contrast, if surface color is \textit{yellow}, the action value is closer to the optimal action value function. As result, the distance optimal action-value on the Bloch sphere is presented in Fig.~\ref{fig:bloch_pretrain}.
\subsection{Bloch Sphere Representation of Local-QNN Pole Training}\label{app:additional_result3}
\begin{figure*}[ht!]
    \centering
\begin{tabular}{cp{0.158\linewidth}p{0.158\linewidth}p{0.158\linewidth}p{0.158\linewidth}p{0.158\linewidth}}
\rotatebox{90}{$\quad 2^{\text{nd}}$ qubit} 
& \includegraphics[width=\linewidth]{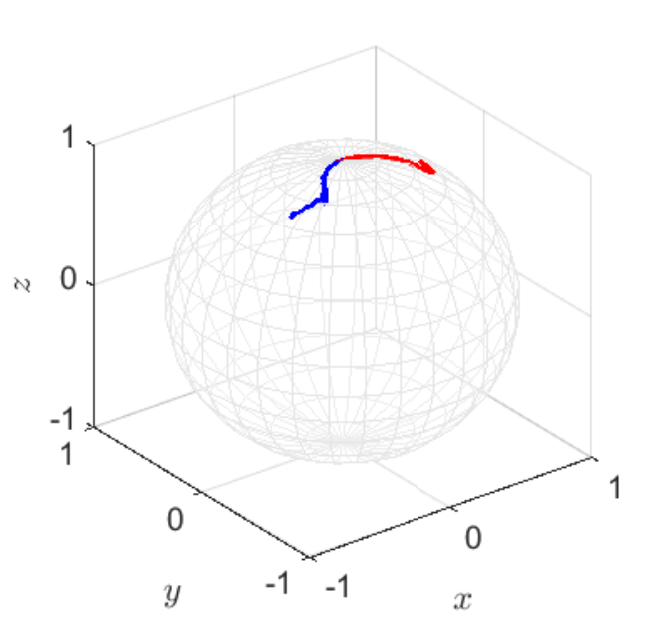}
& \includegraphics[width=\linewidth]{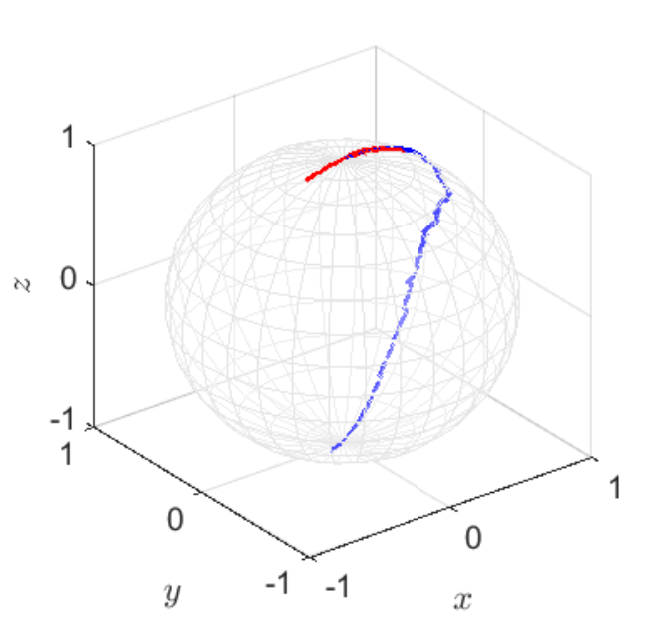}
& \includegraphics[width=\linewidth]{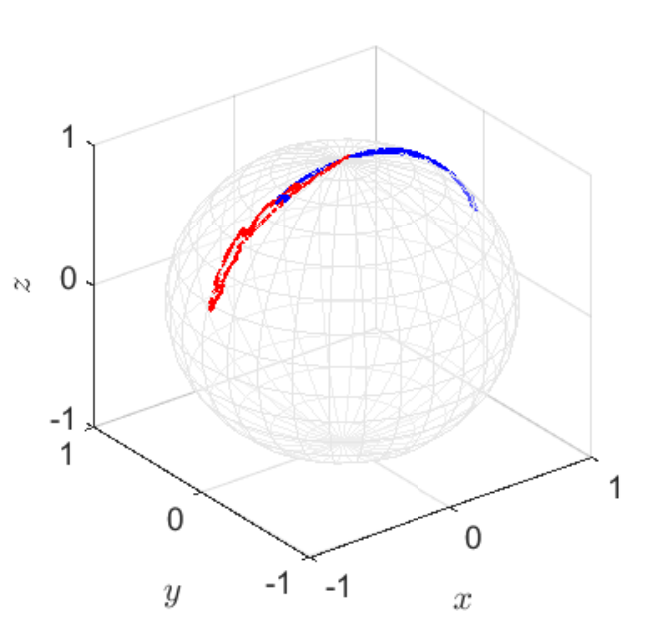}
& \includegraphics[width=\linewidth]{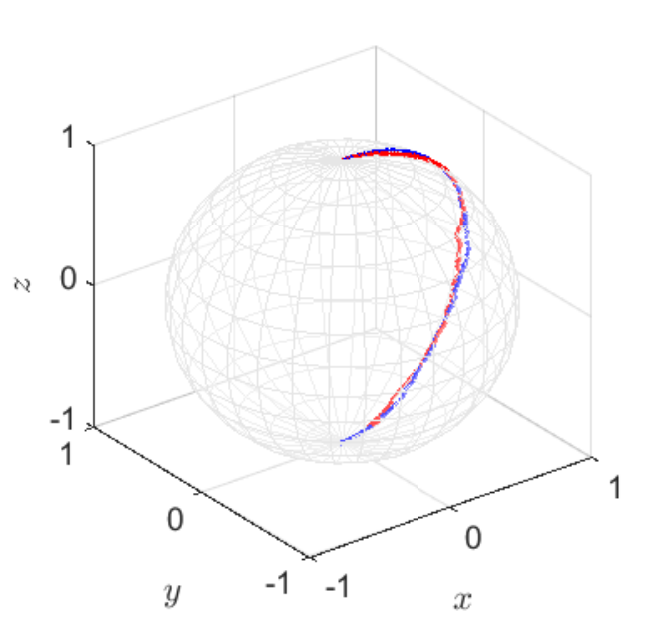}
& \includegraphics[width=\linewidth]{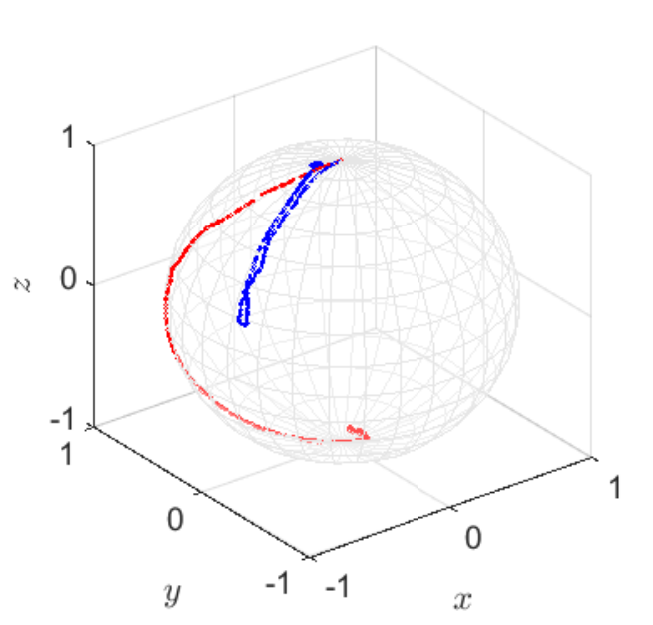}\\
\rotatebox{90}{\quad $3^{\text{rd}}$ qubit} 
& \includegraphics[width=\linewidth]{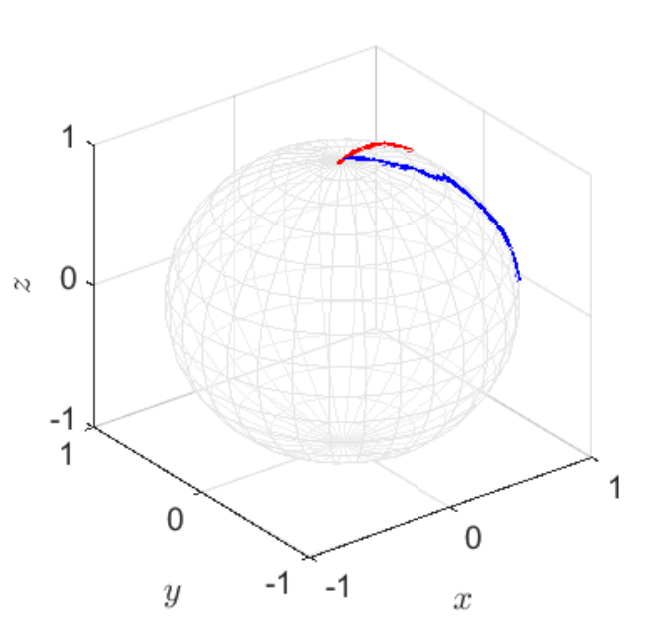}
& \includegraphics[width=\linewidth]{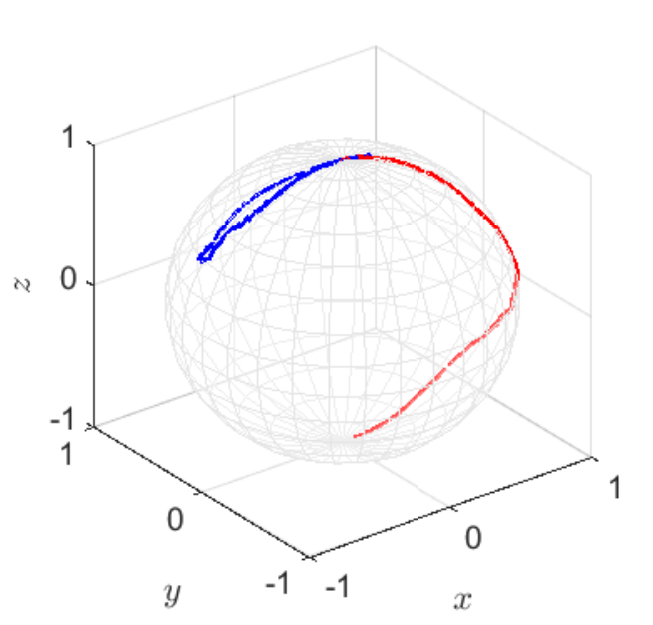}
& \includegraphics[width=\linewidth]{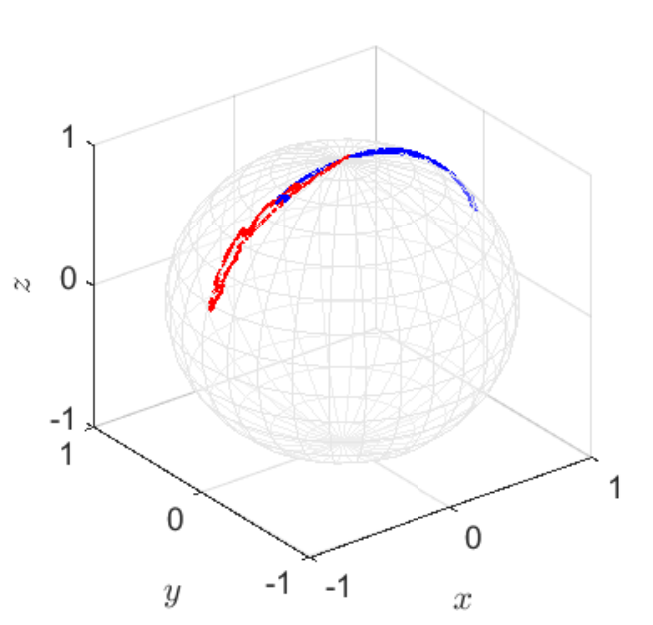}
& \includegraphics[width=\linewidth]{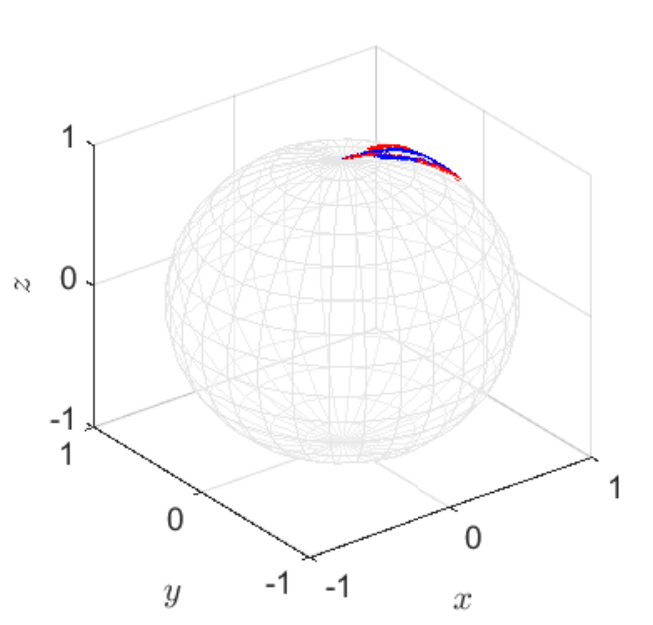}
& \includegraphics[width=\linewidth]{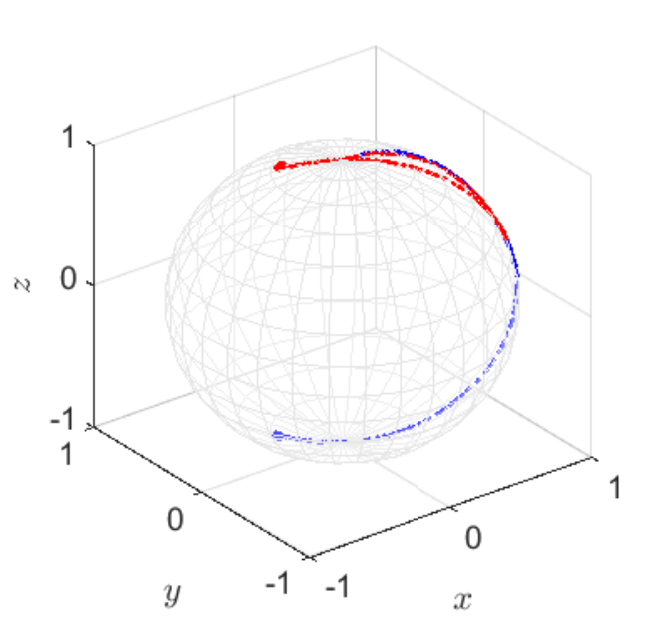}\\
& \multicolumn{1}{c}{\small (a) $\alpha=0^\circ$} 
& \multicolumn{1}{c}{\small (b) $\alpha=30^\circ$} 
& \multicolumn{1}{c}{\small (c) $\alpha=45^\circ$} 
& \multicolumn{1}{c}{\small (d) $\alpha=60^\circ$} 
& \multicolumn{1}{c}{\small (e) $\alpha=90^\circ$}\\
\end{tabular}
    \caption{Tendency of angle training in 3D Bloch sphere; \textit{blue dots} and \textit{red dots} present agent 1 and agent 2, respectively.}
    \label{fig:bloch_measure}
\end{figure*}

We investigate the local-QNN pole training for all qubits associated with the action. The measurement on the second and third qubits are associated to action, \textit{i.e.,} $\mathcal{M}_a = \{2,3\}$. The \textit{blue and red dots} in Fig.~\ref{fig:qvalue} shows the pole of sampled action, \textit{i.e.,} $a \in \arg\max_a Q(s,a;\phis,\thetab)$. We trace all pole parameters associated actions.  Fig.~\ref{fig:bloch_measure} shows the trajectories of agents' measurement parameters on Bloch spheres. 
% \bibliographystyle{unsrtnat}
% \bibliography{references}

\end{document}